\documentclass[twocolumn, twocolappendix, floatfix]{aastex63}
\usepackage{amsmath}
\usepackage{lineno}
\received{\today}
\revised{\today}
\accepted{???}
\submitjournal{ApJ}
\shorttitle{Magnetic fields in the ICM}
\shortauthors{Steinwandel et al.}
\graphicspath{{./}{figures/}}

\begin{document}

\title{Towards cosmological simulations of the  magnetized intracluster medium with resolved Coulomb collision scale\footnote{Released on April, 15th, 2022}}

\correspondingauthor{Ulrich P. Steinwandel}
\email{usteinwandel@flatironinstitute.org}

\author[0000-0001-8867-5026]{Ulrich P. Steinwandel}
\affiliation{Center for Computational Astrophysics, Flatiron Institute, 162 5th Avenue, New York, NY 10010}

\author{Klaus Dolag}
\affiliation{University Observatory Munich, Scheinerstr. 1, D-81679 Munich, Germany}
\affiliation{Max Planck Institute for Astrophysics, Karl-Schwarzschildstr. 1, D-85748, Garching, Germany}

\author[0000-0003-4690-2774]{Ludwig M. B\"oss}
\affiliation{University Observatory Munich, Scheinerstr. 1, D-81679 Munich, Germany}
\affiliation{Excellence Cluster ORIGINS, Boltzmannstr. 2, D-85748, Garching, Germany}

\author{Tirso Marin-Gilabert}
\affiliation{University Observatory Munich, Scheinerstr. 1, D-81679 Munich, Germany}

\begin{abstract}
We present the first results of one extremely high resolution, non-radiative magnetohydrodynamical cosmological zoom-in simulation of a massive cluster with a virial mass M$_\mathrm{vir} = 2.0 \times 10^{15}$ solar masses. We adopt a mass resolution of $4 \times 10^5$ M$_{\odot}$ with a maximum spatial resolution of around 250 pc in the central regions of the cluster. We follow the detailed amplification process in a resolved small-scale turbulent dynamo in the Intracluster medium (ICM) with strong exponential growth until redshift 4, after which the field grows weakly in the adiabatic compression limit until redshift 2. The energy in the field is slightly reduced as the system approaches redshift zero in agreement with adiabatic decompression. The field structure is highly turbulent in the center and shows field reversals on a length scale of a few 10 kpc and an anti-correlation between the radial and angular field components in the central region that is ordered by small-scale turbulent dynamo action. The large-scale field on Mpc scales is almost isotropic, indicating that the structure formation process in massive galaxy cluster formation is suppressing memory of both the initial field configuration and the amplified morphology via the turbulent dynamo in the central regions. We demonstrate that extremely high-resolution simulations of the magnetized ICM are in reach that can resolve the small-scale magnetic field structure which is of major importance for the injection of and transport of cosmic rays in the ICM. This work is a major cornerstone for follow-up studies with an on-the-fly treatment of cosmic rays to model in detail electron-synchrotron and gamma-ray emissions. 
\end{abstract}


\section{Introduction} \label{sec:intro}
Magnetic fields are omnipresent in the Universe and are observed in many astrophysical systems such as compact objects, accretion discs, proto-planetary and proto-stellar discs, the interstellar medium (ISM), planets, stars, and in the largest structures such as galaxies, and the intra-cluster medium (ICM) of galaxy clusters. While magnetic field strengths on the smaller scales in planets, stars compact objects, and accretion discs can reach values from several Gauss (G) to $10^{15}$ G in pulsars, the magnetic fields on the larger scales are, generally speaking, more moderate and typically saturate at the canonical value of a few to a few tens of $\mu$G in galaxies and galaxy clusters but can reach strengths of mG in the dense ISM. \\
Past and current research has developed the following picture of magnetic field amplification in the Universe. A small scale-turbulent dynamo is amplifying tiny seed fields to the values we observe nowadays in galaxies and galaxy clusters at $\mu$G-level. The exact origin of these seed fields is still under debate and several processes have been suggested that can generate seed fields of the order of around $10^{-20}$ G \citep[e.g.][]{Biermann1950, Harrison1970, Demozzi2010, Gnedin2000, Durier2012}. In galaxies, these fields are ordered and further influenced by a large-scale (mean-field) $\alpha$-$\Omega$ dynamo \citep[e.g][]{Parker1955, Steenbeck1966, Parker1979, Ruzmaikin1988} and can be ejected in galactic outflows that can, in turn, magnetize the circumgalactic medium (CGM) \citep[e.g.][]{Bertone2006, Pakmor2017, Pakmor2020, vandeVoort2021}. However, on galaxy cluster scales in the ICM, turbulence driven by the structure formation processes and merger shocks will quickly generate a saturated magnetic field with a field strength of around $\sim \mu$G on the scales of Mpc without the need for an explicit seeding of these fields by galactic winds \citep[e.g.][]{Vazza2018, SteinwandelClusterpaper}. In turn, these fields will then be ordered on larger scales by the structure formation process itself with some evidence that the Void magnetic field can ``remember'' some of the strucutre of the initial seed field on the scales of a few 10 Mpc \citep[e.g.][]{Mtchedlidze2021}.  

\begin{figure*}
    \centering
    \includegraphics[width=0.99\textwidth]{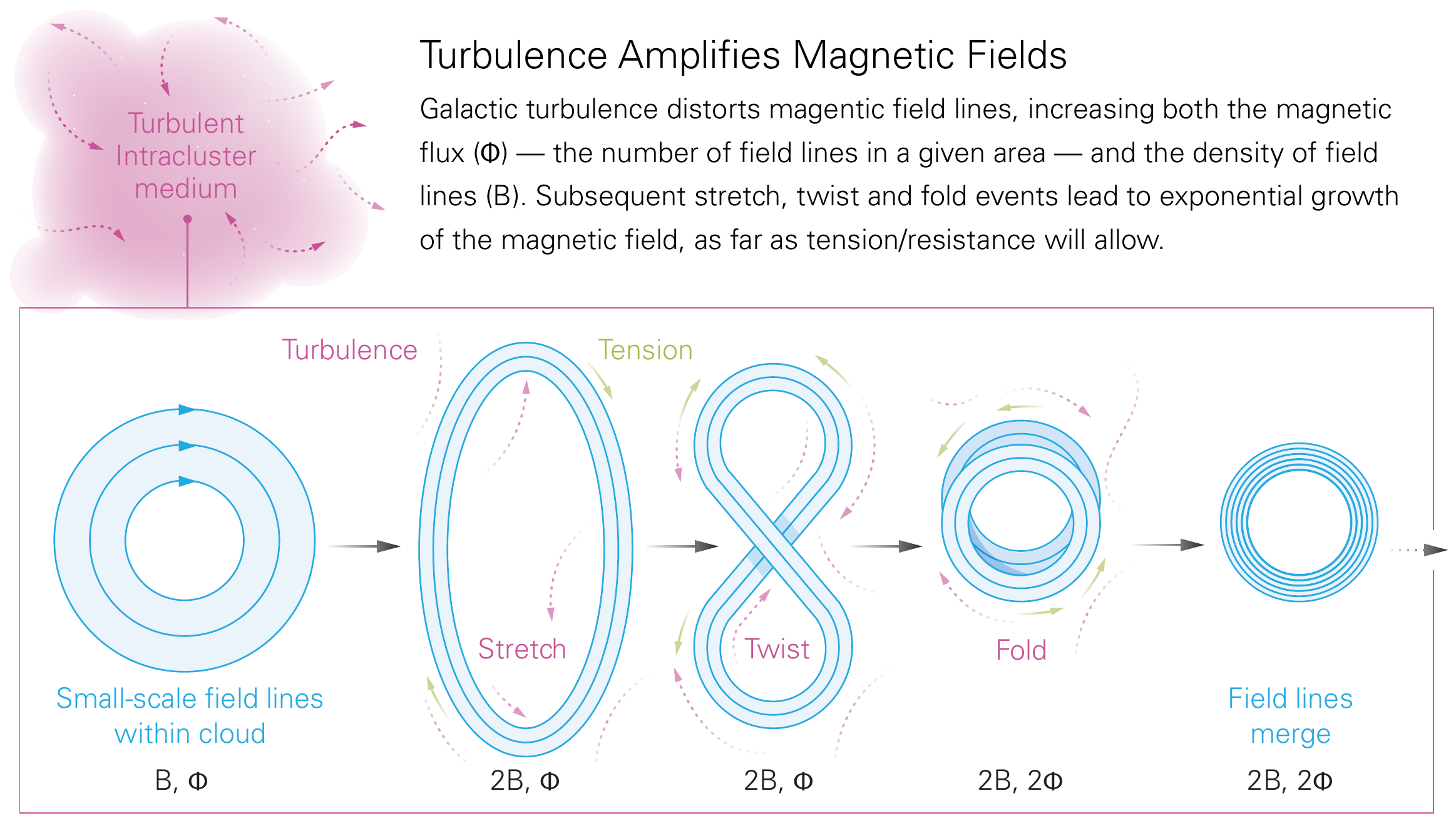}
    \caption{Schematic sketch of the small-scale turbulent dynamo that illustrates the stretching, twisting, and folding of field lines driven by ICM turbulence. As magnetic tension becomes stronger due to subsequent stretching, twisting, and folding, the process slows down and finally saturates when the turbulent kinetic energy in the smallest eddies that drive the process are in equipartition with the magnetic energy density.}
    \label{fig:dynamo}
\end{figure*}

The central process behind the turbulent amplification of magnetic fields is the stretch-twist-fold mechanism as introduced by \citet{Zeldovich1970} but researched by a number of groups \citep[e.g.][]{Kraichnan1967, Kazantsev1968, Kazantsev1985, Kulsrud1992, Brandenburg1995, Kulsrud1997, Xu2020}. The process is schematically described in Fig.~\ref{fig:dynamo}\footnote{We put some effort into this Figure for educational purposes and hope that the community might deem it a useful illustration for the inner workings of the turbulent dynamo in the ICM.}. Small-scale turbulence is first stretching field lines, which is increasing the field strength but at constant magnetic flux. The stretched field lines are then twisted. This step is crucial to note since it requires three spatial dimensions and makes subsequent simulations with reduced dimensions really tricky to interpret. Finally, the twisted field lines are folded, increasing the magnetic flux itself. If this process is repeated, it is easy to understand that it will yield exponential growth in the magnetic field strength. The exponential growth of the field can occur as long as the dynamo stays in the \textit{linear (kinematic) regime} in which the magnetic field is so weak that the tension force of the field lines is much weaker than the force stored in the small scale turbulent eddies. However, as the field strength grows the tension force becomes stronger, and the dynamo transits to the \textit{non-linear regime}, in which the tension force is comparable to the forces exhibited by turbulence. Hence, eventually, the energy stored in turbulence is not enough to perform subsequent folding of field lines and the dynamo \textit{saturates}. Obviously, this depends on the interplay between the amplification of the field and diffusion/dissipation of the field. The magnetic energy is then redistributed from the smaller scales to the larger scales in an inverse cascade. On both galaxy and galaxy cluster scales numerical simulations have well established that this process is dominant in amplifying magnetic fields using different numerical prescriptions \citep[e.g.][]{Dolag1999,Dolag2002,Kotarba2009,Dubois2008, Wang2009, Pakmor2013, Pakmor2017, Butsky2017, Garaldi2020, Steinwandel2019, Steinwandel2020, Steinwandel2022, SteinwandelClusterpaper, Vazza2014, Vazza2018}. However, small-scale dynamos can only generate correlated fields on the scale of the turbulence and require a process that is ordering the field on the largest scales. For instance, one can derive the outer scale of MHD turbulence by using the peak in the magnetic energy spectra that is essentially set by the magnetic Reynolds number that compares the advection time scale with the magnetic  diffusion time scale. The exact interplay between those two timescales will set the peak of the magnetic power spectra and thus the outer scale of the underlying MHD turbulence. In other words, the scale of equipartition is set by the peak of the underlying magnetic power spectrum and the exact nature of MHD turbulence. 

In this paper, we present the first results from a modern state-of-the-art galaxy cluster formation simulation that is specifically targeted to understand the complex properties of the ICM in a fully cosmological context at unprecedented resolution. Hereby, we will, for the first time demonstrate that the high resolution targeted in this paper marks the endpoint for a classic MHD treatment on galaxy cluster simulations as we will show that the coulomb mean free path is resolved over a vast regime of typical densities and temperatures in the ICM.

While there is some progress in the modeling of low-mass galaxy clusters \citep[e.g.][]{Kannan2017, Tremmel2019, Pillepich2019, Ricarte2021, Butsky2019} massive galaxy clusters are only rarely studied in cosmological zoom simulations or large cosmological volume simulations. The reason for this is twofold. First, these objects often follow more complex accretion scenarios than lower-mass objects such as low-mass galaxy clusters ($\sim$ 10$^{14}$ M$_{\odot}$) and galaxy groups ($\sim$ 10$^{13}$ M$_{\odot}$) that can include several (major and minor) mergers of the latter objects. Second, the ICM on large scales is governed by plasma-astrophysical processes such as magnetic fields \citep[e.g.,][]{Drake2021, Berlok2022, Squire2023, Kunz2016, Kunz2019, Kunz2022, Vazza2014, Vazza2018, SteinwandelClusterpaper}, (anisotropic) viscosity \citep[e.g.,][]{Sijacki2006, Berlok2020, Marin2022}, and (anisotropic) conduction \citep[e.g.,][]{Kannan2017, Hopkins2017, Berlok2021} as well as cosmic rays \citep[e.g.,][]{Sijacki2008, Boess2023}, that can significantly contribute to its phase structure. Hence, these processes have to be modeled appropriately in simulations of more massive galaxy clusters, which is mostly done in detailed ``turbulent driven studies'' \citep[e.g.,][]{Schekochihin2004,Porter2015, Mohapatra2021, Mohapatra2022} and only rarely in large scale fully cosmological simulations. The complications when it comes to simulating such systems while including all the relevant plasma astrophysical processes are obvious. The numerical treatment of these processes is not only computationally expensive but requires also very high resolution to resolve the complex structure of the ICM. Moreover, the ultimate goal should be to push massive galaxy cluster simulations in a regime where they can start to capture the effects of smaller scale instabilities, such as the magneto thermal instability \citep[MTI; e.g.,][]{McCourt2012} or the heat flux driven buoyancy instability \citep[HBI; e.g.,][]{Parrish2008}, for which the presented simulation can be an important cornerstone to achieve this goal. In order to understand the detailed impact of such instabilities on the structure of the ICM, one first needs to understand fundamental plasma astrophysical mechanisms such as magnetic field amplification, and gauge that the a priori amplification of small seed fields provides the conditions that are necessary to produce the high $\beta$ plasma needed for such  instabilities to form. Furthermore, it remains unclear how important such instabilities remain for the thermal state of the ICM because most of the simulations carried out where these effects arise are much more idealized than a fully cosmological simulation. Hence one needs to work continuously towards higher resolution, fully cosmological simulations with the appropriate plasma astrophysical treatment in order to resolve the scale on which plasma instabilities can build up. The simulation presented in this paper represents an important link to achieving this goal with massive galaxy cluster simulations.\\
In this paper, we present the first results of a simulation of  a massive galaxy cluster with a total mass of $2 \times 10^{15}$ M$_{\odot}$ modeled with the full treatment for magnetohydrodynamics (MHD). We will study the amplification of the field at a resolution that is high enough to resolve the Coulomb collision scale in the ICM, and provide important insights into the effect of magnetic fields on the thermal pressure profiles of galaxy clusters. 

\begin{figure*}
    \centering
    \includegraphics[width=0.49\textwidth]{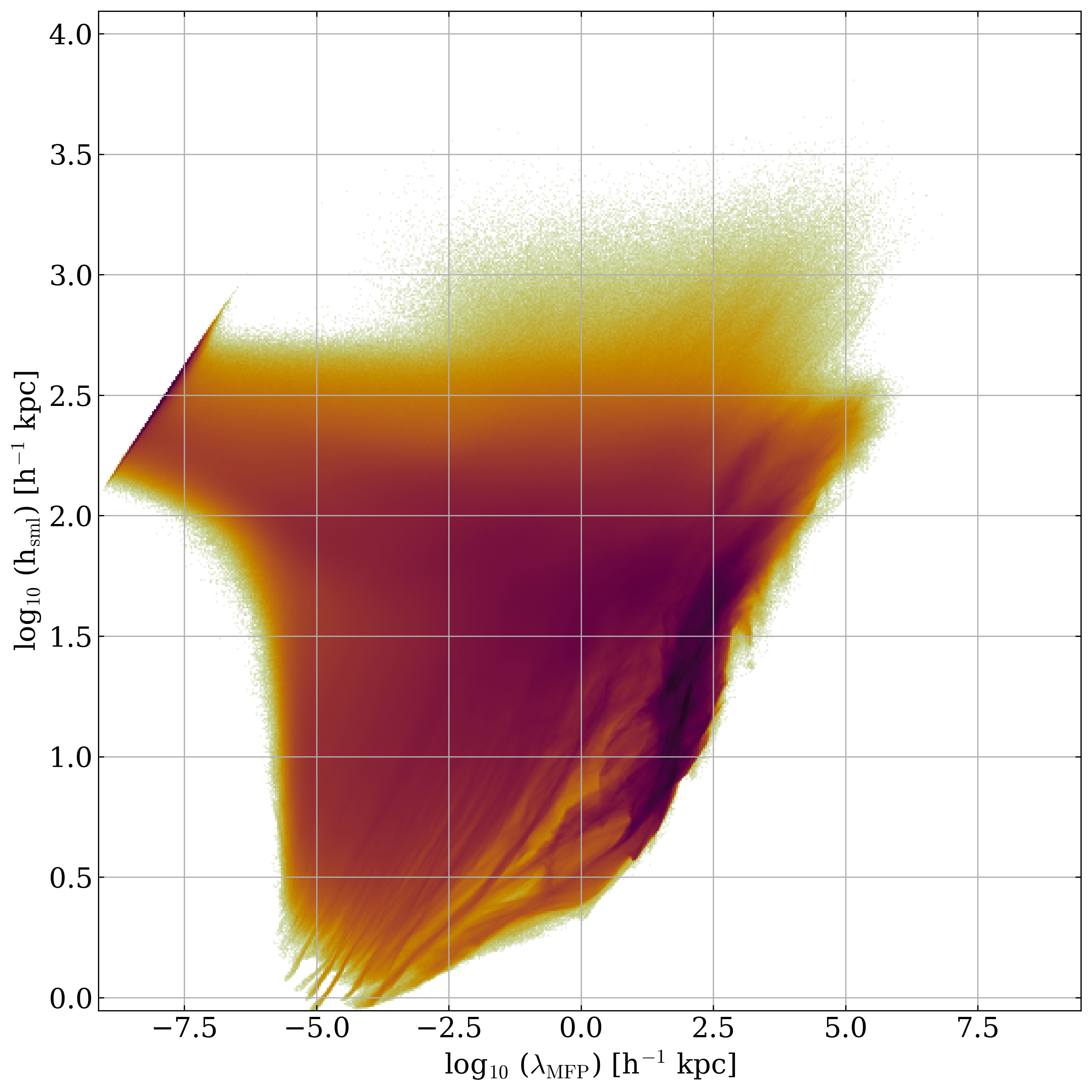}
    \includegraphics[width=0.49\textwidth]{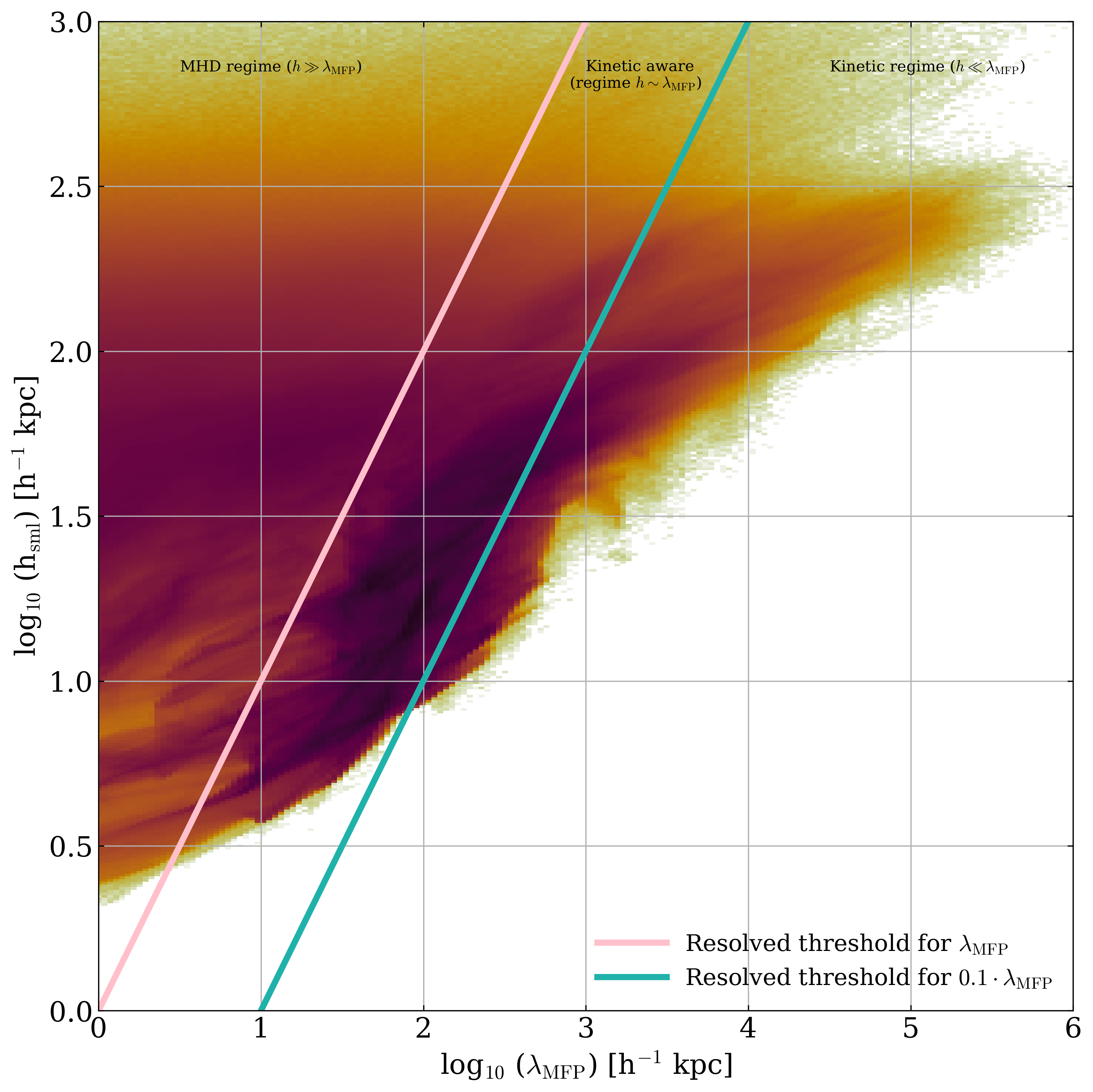}
    \caption{Left: Global distribution of the resolution as a function of the electron mean free path. Right: Resolution as a function of the electron mean free path as a measure for the Coulomb diffusion length scale. We show this for the region in the ICM that caries most of the mass in the ICM between $10^{-7}$ cm$^{-3}$ and temperatures above $10^6$ K (see Fig.~\ref{fig:phase_temp}). It is apparent that our resolution in this regime resolves the Coulomb mean free path ($\lambda_{MFP}$) over a broad density and temperature regime in the ICM. The oink and turquoise line mark the transition regions at which the smoothing length h$\approx \lambda_\mathrm{MFP}$ and h$\approx 0.1 \times \lambda_\mathrm{MFP}$ indicating that we transit into a regime where kinetic aware MHD becomes an interesting addition. To the right of the turquoise line, one should consider a fully kinetic treatment. This indicates that our current resolution marks the limit for an MHD-only treatment for galaxy cluster simulations.}
    \label{fig:mean_free_path}
\end{figure*}

\begin{figure*}
    \centering
    \includegraphics[width=0.99\textwidth]{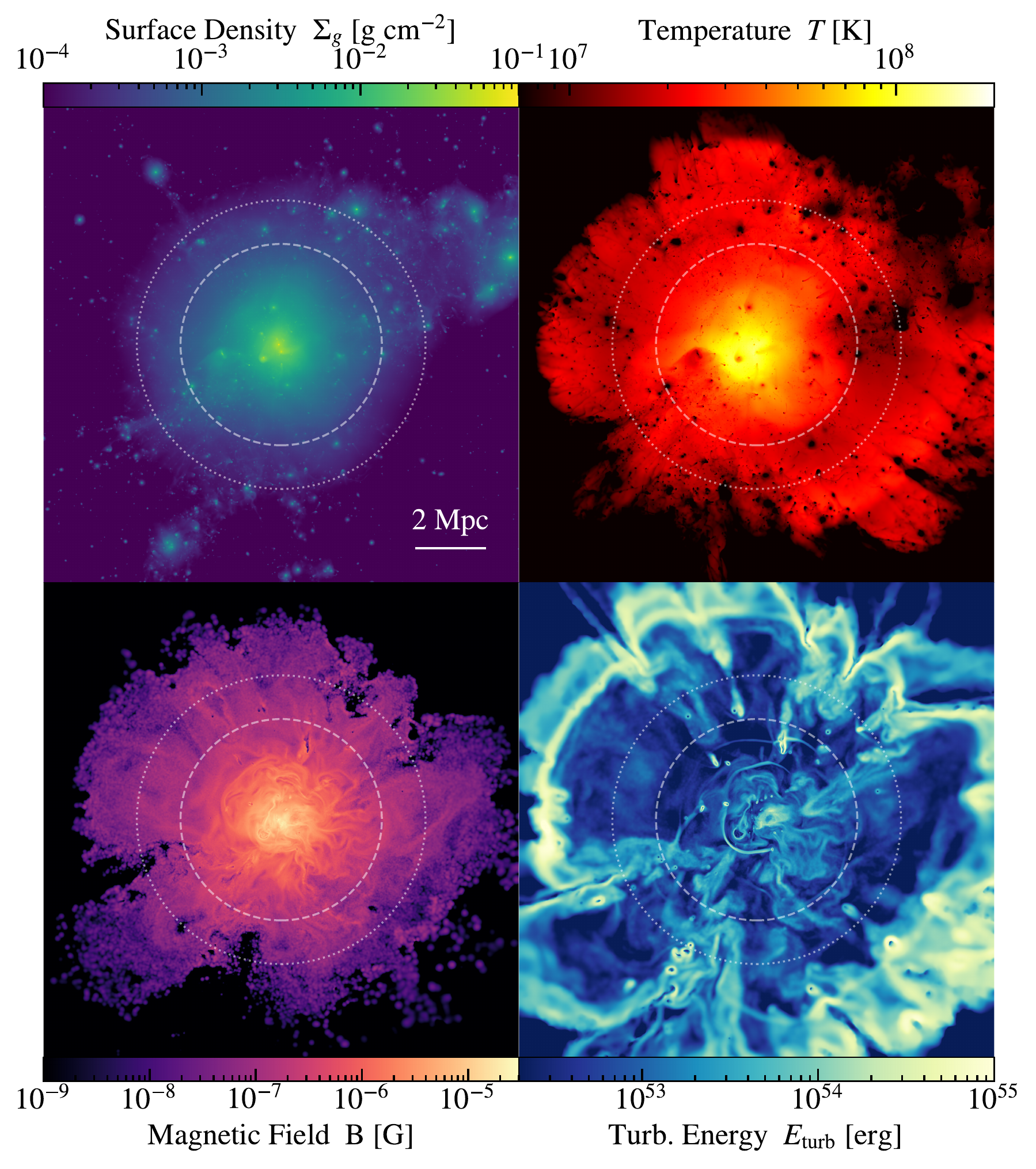}
    \caption{Surface density (top left, projection), the temperature (top right, projection), the magnetic field strength (bottom left, slice with 100 kpc height) as well as the turbulent energy (bottom right, slice with 100 kpc height) for the simulation at redshift 0. The magnetic field at redshift $z=0$ is in rough agreement with observed values in the Coma cluster as reported for example by \citep[e.g.,][]{Bonafede2010}. However, we note that the field is slightly higher than in observations (factor of three). Similar to previous galaxy cluster simulations the magnetic field drops quickly as a function of the radius. The turbulent energy reveals the complex shock structure we find in the ICM that is dominated in the center by internal weakly supersonic shocks and in the outskirts dominated by the strongly supersonic accretion shock. The dotted and dashed white circles indicate R$_{200}$ and R$_{500}$ respectively.}
    \label{fig:cluster}
\end{figure*}

\begin{figure*}
    \centering
    \includegraphics[width=.99\textwidth]{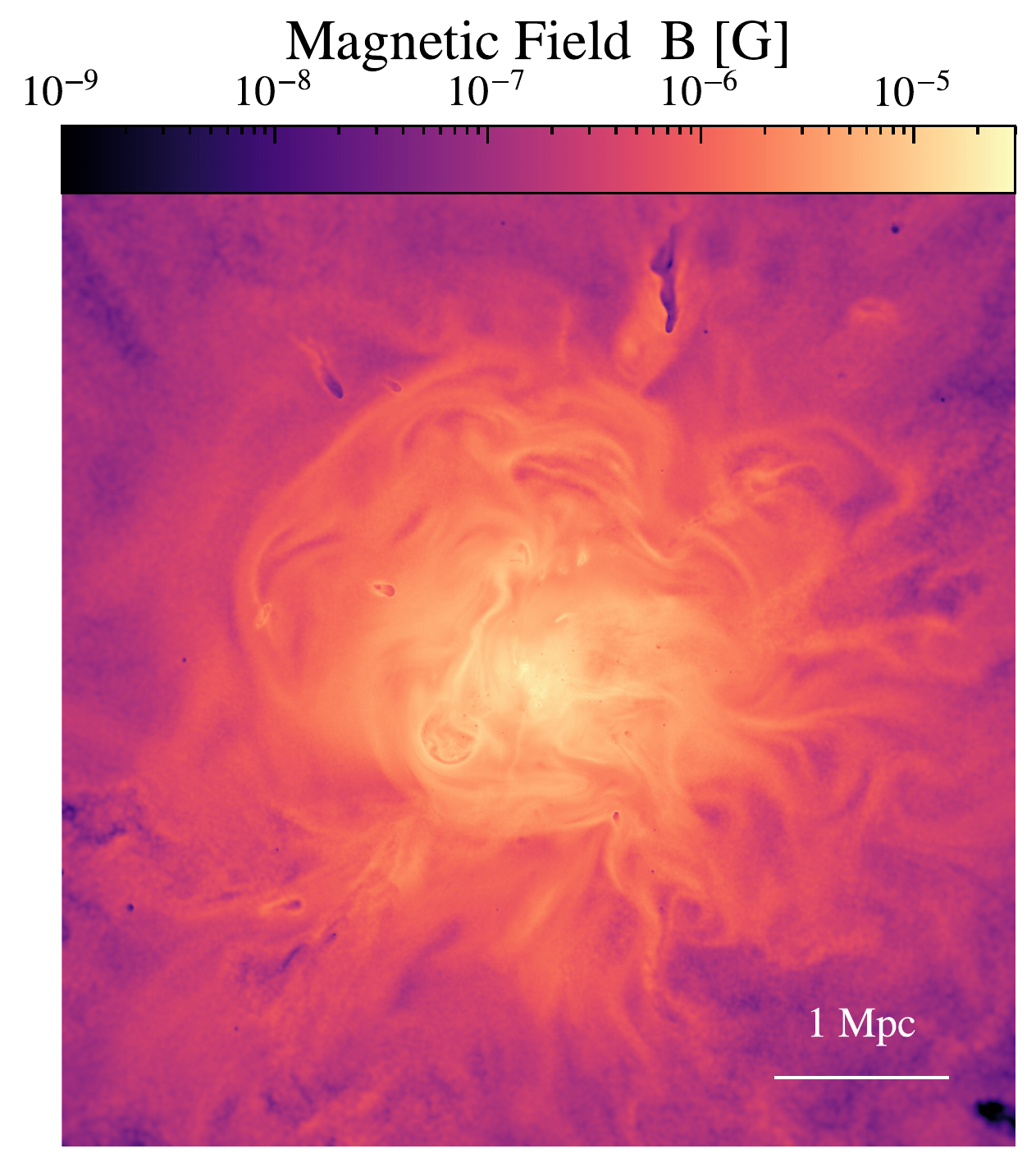}
    \caption{Central structure of the magnetic field in a very thin slice of $\Delta x =100$ kpc alongside the xy-direction of the cluster. We choose this orientation to demonstrate how the small and large-scale field structure are related to one another. The magnetic field in the center reaches a magnitude of around 20 $\mu$G on the smallest scales. The magnetic field is highly turbulent and not correlated. However, there are regions that show large-scale orientation on Mpc scales that likely originate from shocked gas. This indicates that the field is ordered on larger scales by the large-scale structure formation process.}
    \label{fig:cluster_yz}
\end{figure*}

\begin{figure*}
    \centering
    \includegraphics[width=0.49\textwidth]{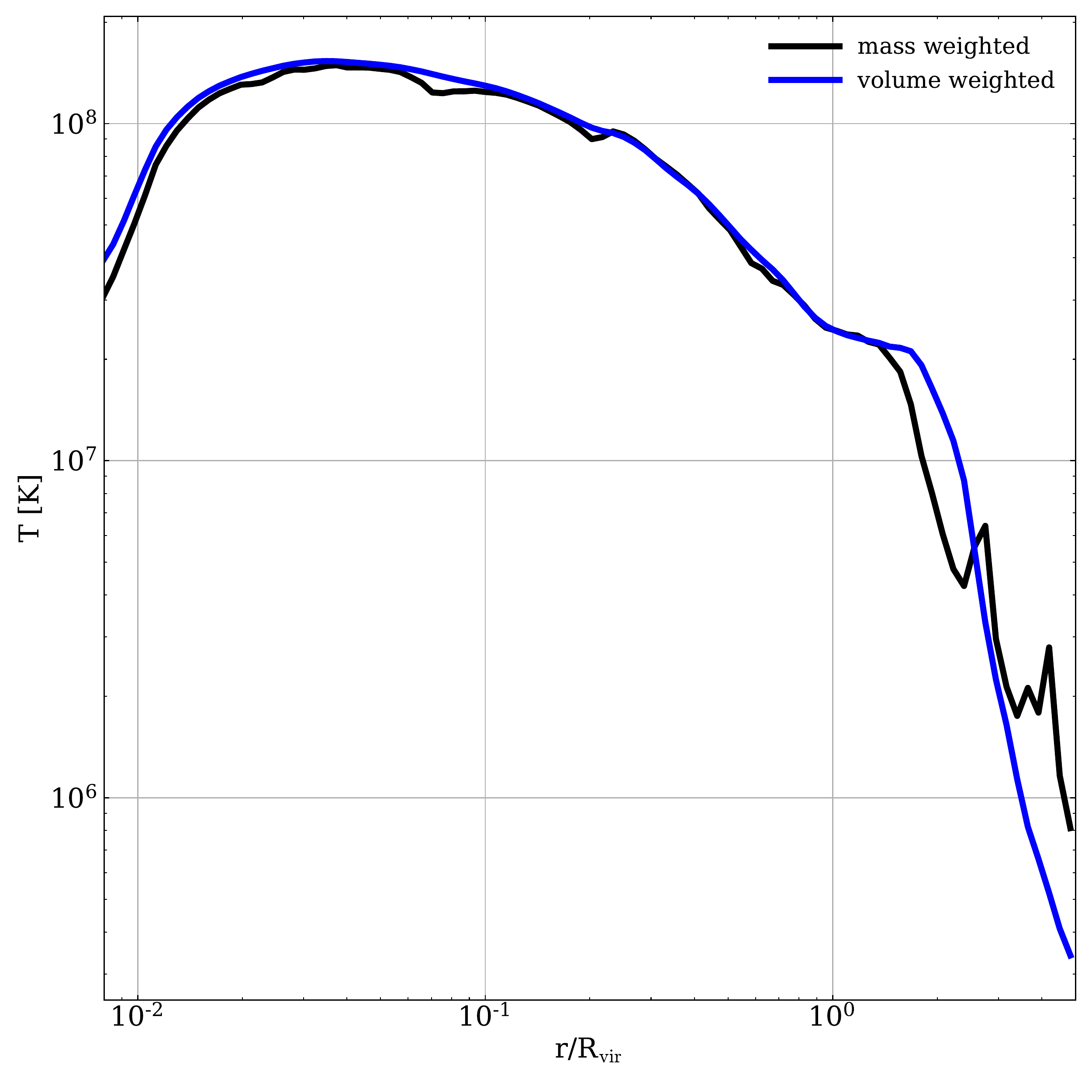}
    \includegraphics[width=0.49\textwidth]{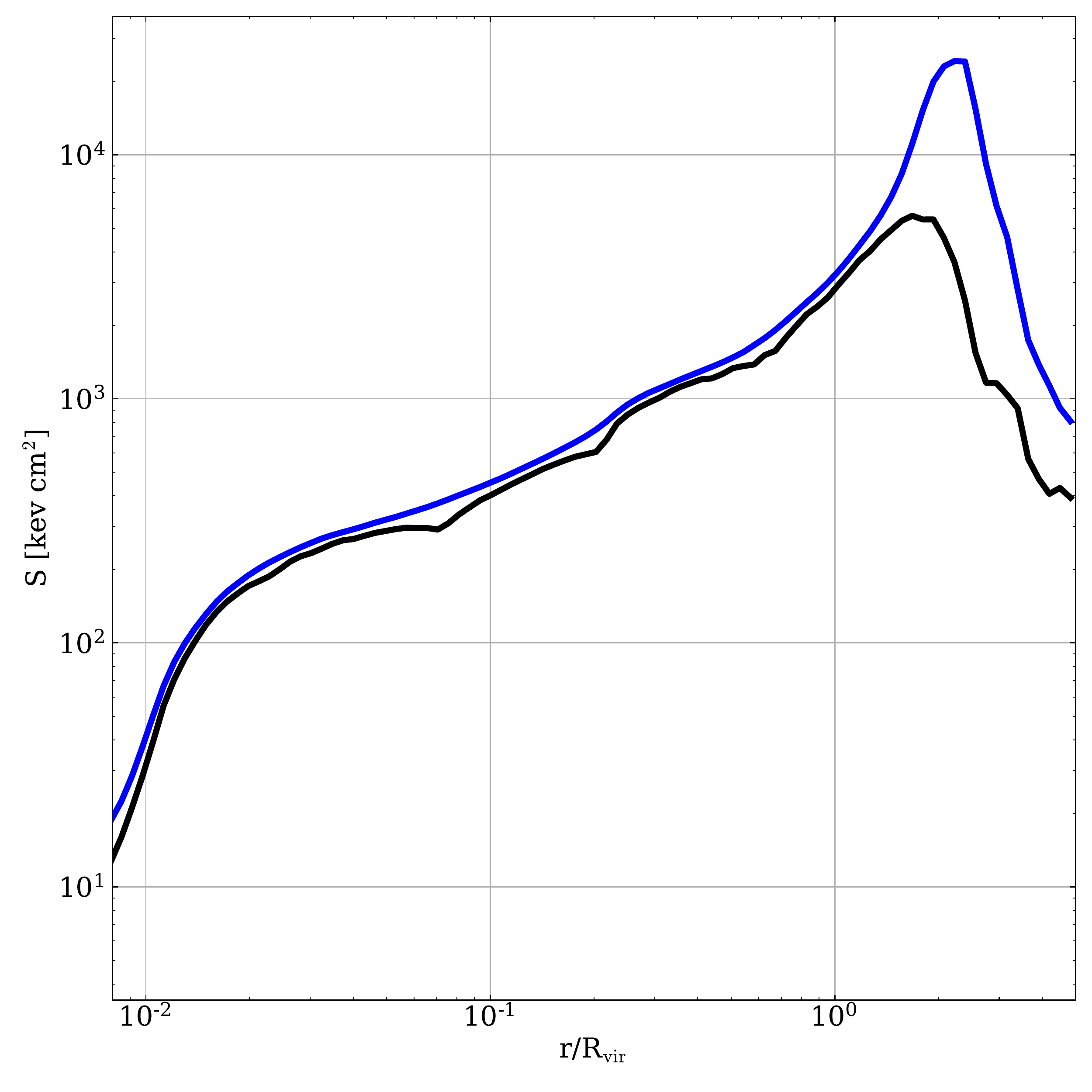}
    \includegraphics[width=0.49\textwidth]{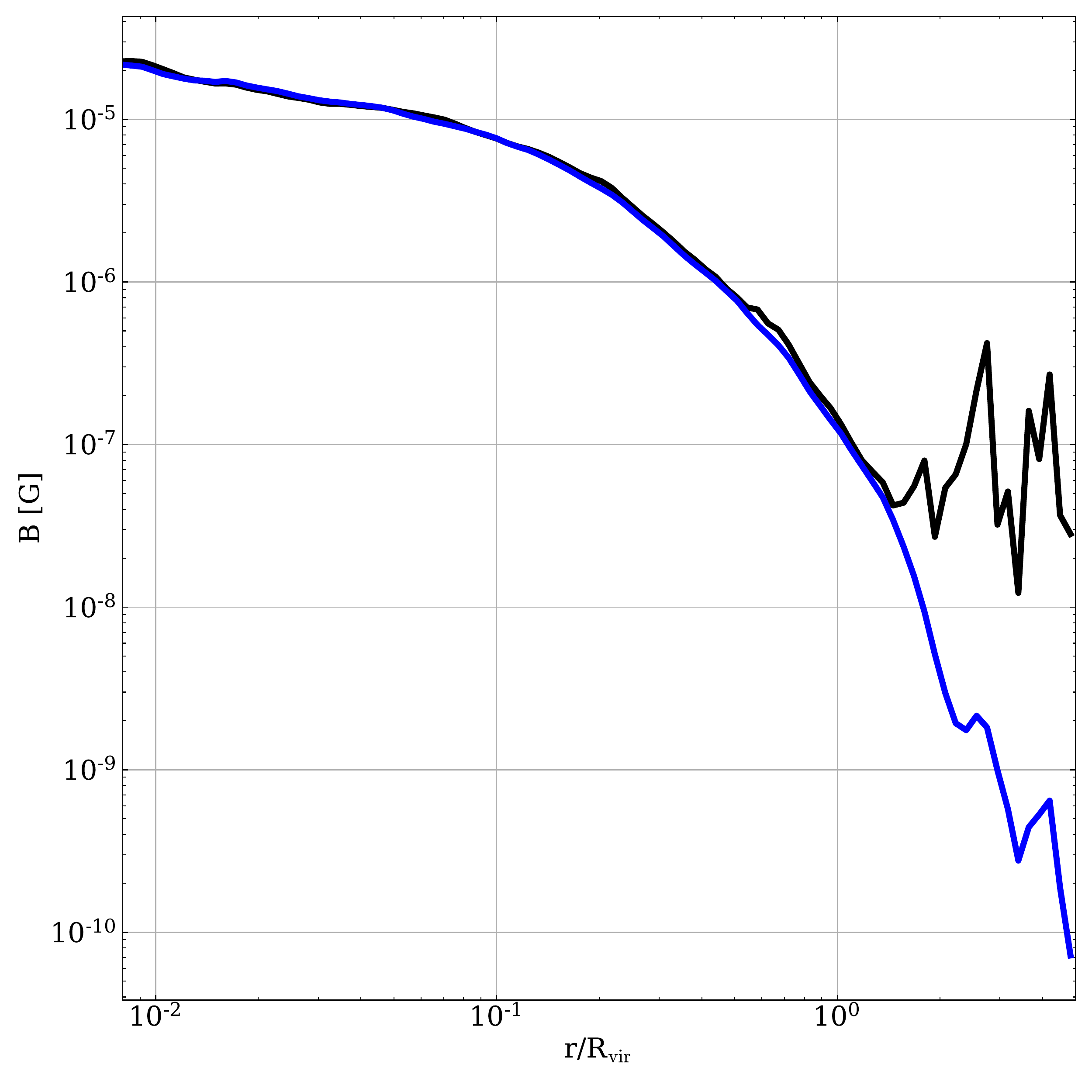}
    \includegraphics[width=0.49\textwidth]{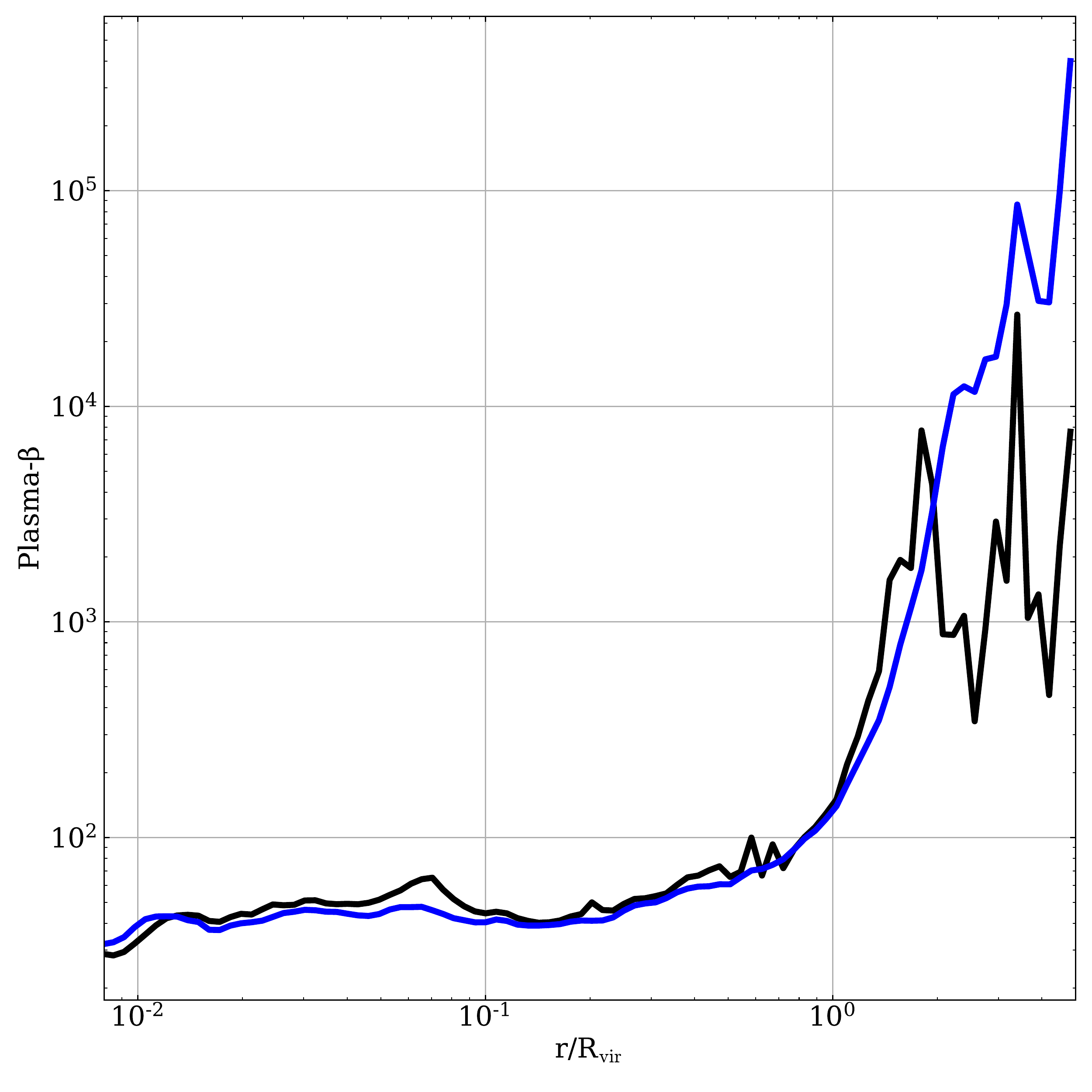}
    \caption{Radial profiles of the magnetic field (top left), the Plasma$\beta$ parameter (top right), the temperature (bottom left), and the entropy (bottom right) for the simulated system at redshift zero. The cluster shows well-behaved ICM properties with magnetic field strengths ranging around $10$ $\mu$G. The values of Plasma-$\beta$ are high in the center with values between 40 and 100 but decrease quickly in the outskirts below to a few 100, which is in good agreement with the general belief that the ICM is high-$\beta$ plasma. Temperatures in the center are high between $10^7$ and $10^8$ K. The entropy profile drops strongly in the center due to a steep drop in temperature.}
    \label{fig:radial_profiles}
\end{figure*}

\section{Numerical Methods} \label{sec:numerics}

\subsection{Simulation code}

We briefly present the numerical methods used and the numerical simulations discussed in this paper. We use the Tree-SPMHD code \textsc{gadget-3} to carry out all the simulations presented in this work. Gravity is solved via the Tree-PM method where the long-range gravitational forces are computed on a PM mesh and the short-range forces are computed on the gravity tree. This reduces the workload on the tree significantly, most notably in terms of the memory imprint of the code (and a factor of around 2 in total run time for a given simulation, although this depends on the setup). Furthermore, the code has the option to use a split PM-mesh for zoom initial conditions that can have arbitrary resolution compared to the large-scale PM-mesh. However, for this simulation, we disabled the option of using a second PM grid and the force computation is somewhat more accurate as only the tree is used for updating forces on most of the high-resolution zoom region. Furthermore, truncation errors are suppressed in the zoom region as there is no need for interpolating between the forces computed by the gravity tree and the PM algorithm. \\
The code utilizes a modern prescription for SPH that includes higher order kernels \citep[e.g.,][]{Wendland1995, Wendland2005, Dehnen2012} and a treatment that leads to an  improved mixing behavior of SPH in shear flows based on artificial viscosity and artificial conduction \citep[e.g.][]{Price2012, Hopkins2013, Hu2014, Beck2016} following the implementation of \citet{Beck2016}. \\ 
Magnetohydrodynamics (MHD) is introduced based on the implementation of \citet{Dolag2009} with the updates of \citet{Bonafede2011} that includes a treatment for non-ideal MHD. The non-ideal MHD is handled over a constant (physical) diffusion and dissipation. For the latter, we heat the gas with the magnetic field that is lost due to magnetic reconnection. We model (an)isotropic conduction via a conjugate gradient solver \citep{Petkova2009, SteinwandelSNpaper}, that has been extended towards a bi-conjugate gradient solver in \citet{Arth2014}, and \citet{SteinwandelClusterpaper} with the specific use case for massive galaxy cluster formation simulations that include magnetic fields. The simulation in this paper is carried out with a suppression factor of the Spitzer value for conduction of 5 per cent \citep[e.g.,][]{Spitzer1953, Spitzer1956}. We adopt a Wendland C4 kernel with 200 neighbors and bias correction as suggested by \citet{Dehnen2012}.

\subsection{Initial Conditions and Simulations} \label{sec:ICS}

We present the results of one high mass galaxy cluster zoom simulations at an unprecedented resolution of a halo with a total mass of M$_\mathrm{tot} = 2 \times 10^{15}$ M$_{\odot}$ where we reach a spatial resolution of around 0.250 kpc and a mass resolution of a $4 \times 10^{5}$ M$_{\odot}$ in the ICM. The simulation ``250X-MHD'' has been performed in New York on the in-house cluster ``rusty'' of the Simons Foundation and consumed around 6 million core hours (excluding halo finding and debug runs). The simulation was performed in a non-radiative fashion without cooling and star formation. Hence, it is more targeted for understanding the complex plasma physical aspects of the ICM, rather than the galaxy formation process in super-massive clusters. The reason for this is two-fold. First, we want to understand how magnetic fields can contribute to the structure formation process on the largest scales in a clean experiment that is not dominated by underlying subgrid prescriptions for star formation and feedback. Second, a full physics run for one of these clusters that include treatment for cooling, star formation, and the feedback of stars and active galactic nuclei (AGN) is $\sim 10$ times more expensive. However, once we have a better understanding of the plasma astrophysical aspects of the ICM in a fully cosmological simulation, we will attempt this simulation with a full feedback prescription but will postpone the results for future work.\\

The initial conditions for the cluster are chosen from a lower-resolution dark matter-only simulation of a Gpc volume \citep{Bonafede2011}. Only in Gpc volume one can find a sample of massive clusters as the one re-simulated here in abundance. The base dark matter simulation has a resolution of 1024$^3$ particles leading to an overall mass resolution of 10$^{10}$ M$_{\odot}$. The cosmological parameters for the simulation are chosen based on WMAP7 cosmology with $\Omega_{0} = 0.24$, $\Omega_{\Lambda} = 0.76$, $\Omega_\mathrm{baryon} = 0.04$, $h = 0.72$ and $\sigma_{8} = 0.8$. We select dark matter particles at $z=0$ for one of the most massive halos in the box and trace them back with the method described in \citet{Tormen1997} to obtain zoomed initial conditions. This cluster has been previously simulated with hydrodynamics only in \citet{Zhang2020a, Zhang2020b}. 

The domain from which we start re-simulation is large enough to avoid massive intruder particles within 5 times the virial radius at redshift zero. The magnetic field is initialized as a constant comoving seed field of 10$^{-14}$ G. This choice marks a quite large seed field and leads to a saturated dynamo by redshift 2. We tested this in detail in our lower resolution versions of this cluster in \citet{SteinwandelClusterpaper} and noted that a change of this seed field by a factor of 10 (lower or higher) produces similar results (although a lower seed field in combination with a higher magnetic diffusivity yielded lower mean magnetic fields in radial profiles). We note that for arbitrarily small seed field (values below 10$^{-20}$ G) we find no saturated dynamo in runs \textit{without} galaxy formation physics (cooling, star formation, stellar- and AGN-feedback). However, we note that we only tested this for the lowest resolution version presented in \citet{SteinwandelClusterpaper} and this could obviously change in the higher resolution versions of this cluster of which we currently run a few micro-physics variations. However, for now, we just focus on our fiducial MHD run and specifically the detailed structure of the magnetic field itself.

\section{Results} \label{sec:results}

In this section, we will present our results and describe the major plots that are important for the study.

\subsection{Resolving the electron mean free path}

We start the presentation of our results with Fig.~\ref{fig:mean_free_path} where we show the spatial resolution of our SPH simulation (smoothing length) as a function of the electron mean free path, color-coded by mass (we note that we omit the color bars since this is 2d joint PDF). The pink and the turquoise line mark the regimes in which $\lambda_\mathrm{MFP}$ and $0.1 \times \lambda_\mathrm{MFP}$ are marginally resolved. We compute the mean free path following \citet{Zeldovich1967}:
\begin{align}
    \frac{1}{\lambda_\mathrm{MFP}} = \frac{2 \pi}{9} \cdot n_\mathrm{e} \frac{Z^4 e^4}{k_\mathrm{B}^2 T^2} \ln \Lambda,  
\end{align}
where $n_\mathrm{e}$ is the electron number density, and T is the temperature of each particle in the simulation. Z is the atomic number for which we adopt one (electrons, protons) and $e$ is the elementary charge of $4.803^{-10}$ statC (again electrons/protons). For $\ln \Lambda$ we adopt 30, which seems to be in good agreement with typical values in the ICM. We find that the bulk of the mass is located between $10^{-7}$ cm$^{-3}$ and $10^{-2}$ cm$^{-3}$ and is therefore lying in between the bounding lines for resolved $\lambda_\mathrm{MFP}$ and $0.1 \times \lambda_\mathrm{MFP}$. This indicates that our simulation operates at the limit where the classic MHD equations are a good approximation. Hence, future simulations at our resolution need to investigate the inclusion of anisotropic viscosity and heat conduction. Simulations at 10 times higher mass resolution should push for a more sophisticated framework that includes a closure for the detailed plasma kinetics when the resolution of the MHD simulation becomes much smaller than the mean free path of the electrons (and ions respectively). We will discuss the implications of this in greater detail in section \ref{sec:mean_free_path}.

\subsection{Structure and Morphology}
In Fig.~\ref{fig:cluster} we show the MHD simulation at redshift zero. The panels describe the gas surface density (top left), the temperature (top right ), the structure of the magnetic field strength (bottom left), as well as the turbulent kinetic energy (bottom right). The white circles indicate R$_{200}$ (dotted) and R$_{500}$ (dashed). The virial radius of the system is $\sim 2.3$ Mpc h$^{-1}$ at redshift zero. All the panels are obtained by binning the data to a uniform rectangular grid, using an SPH interpolation to the cell center, based on a slice in z-direction centered 100 kpc around the clusters' x and y position. Generally, we note that the cluster is a very relaxed system at redshift zero (there are no major mergers happening at this redshift) but shows a rather complex history of merger events with several significant major merger processes in its formation history. The most significant ones are at redshifts of $z=0.3$, $z=0.8$, $z=1.2$, and $z=2.7$ and are discussed in more detail for the lower resolution versions of this system in \citet{SteinwandelClusterpaper}. The density projection reveals a lot of substructure beyond the virial radius falling into the cluster center, even at redshift $z=0$, that will lead to subsequent major merger events in the near future. For instance, the second most massive halo in the simulation has still a mass of around $10^{14}$ M$_{\odot}$ and its outskirts can be seen at the top right of the top left panel of Fig.~\ref{fig:cluster}.  
The temperature distribution shows peaks of a few times $10^{8}$ K. The magnetic field structure at $z=0$ is fully developed and saturates at the level of a few $\mu$G, but shows significantly higher values at larger redshift (not discussed in detail in this paper), which is consistent with earlier galaxy cluster simulations and the higher RM-values, typically observed under high redshift conditions. Additionally, the structure in the turbulent kinetic energy reveals the detailed structure of both internal MHD shocks as well as the external high Mach number accretion shock located beyond the virial radius of the shock. It is noteworthy to point out the incisiveness of the detailed MHD cluster shocks.  

In Fig.~\ref{fig:cluster_yz} we show a very thin slice of the xy-direction to illustrate the turbulent, highly uncorrelated but fully developed magnetic field structure, especially within the virial radius. The complex field structure observed in the virial radius indicates that the field is organized on larger scales by the large-scale structure formation. It is quite clear from the visualization that the magnetic field has a typical correlation length of several 100 kpc. This is larger than the typical size of the smallest turbulent eddies in the simulation with characteristic sizes of around 1 kpc. 

In Fig.~\ref{fig:radial_profiles} we show the radial evolution of central plasma parameters such as the magnetic field strength (upper left panel), Plasma-$\beta$ (upper right panel), the temperature (bottom left panel) and the entropy (bottom right panel) mass (black) and volume-weighted (blue). We find central magnetic field strengths of up to 20 $\mu$G comparable to those reported in our previous work \citet{SteinwandelClusterpaper} for our lower resolution runs 25X-MHD and 10X-MHD. 

The Plasma-$\beta$ parameter is typically high with central values of around 50 which increases to around 200 when approaching the virial radius of the system. Beyond the virial radius, Plasma-$\beta$ is generally high between 100 and a few times 1000. While the values in the center are low, they are still well above unity and the magnetic field remains dynamically unimportant for the formation process of the cluster. However, we note that obviously, the field structure will be important for cosmic ray injection, re-acceleration, and propagation. To that end, we developed the spectral cosmic-ray model described in \citet{Boess2023} for electrons and protons which we will apply to a sister simulation on the fly in future work.   

The radial temperature profiles indicate that the ICM has a typical temperature of $10^{8}$ K ($\approx 10$ keV), which drops towards the center of the cluster by roughly a factor of two ($\approx 4$ keV). The decrease of temperature in the center is in direct relation to the rather high Plasma-$\beta$ values of around 50 that we find in the simulation. The reason for the decrease in temperature is likely influenced by the absence of important feedback physics that would increase entropy in the cluster center. We show our entropy profiles in the bottom right panel of Fig.~\ref{fig:radial_profiles} that bottom out towards the center. That in turn leads to a loss of thermal support of the cluster and a decrease of Plasma-$\beta$ towards the center. As mentioned above, the decline in the central entropy profile is supported by the absence of heating by AGN-feedback. We do note that some of our simulations with a higher thermal conductivity than our adopted suppression coefficient of 0.05 show flat entropy cores around $10^{3}$. This topic will be the subject of more detailed future studies. Moreover, we point out that the peak in the entropy profile past the virial radius marks the accretion (virial shock) of the  cluster. Hereby it is interesting to point out that the volume-weighted prescription gives a more accurate position of the virial shock than the mass-weighted prescription. 

\begin{figure}
    \centering
    \includegraphics[width=0.49\textwidth]{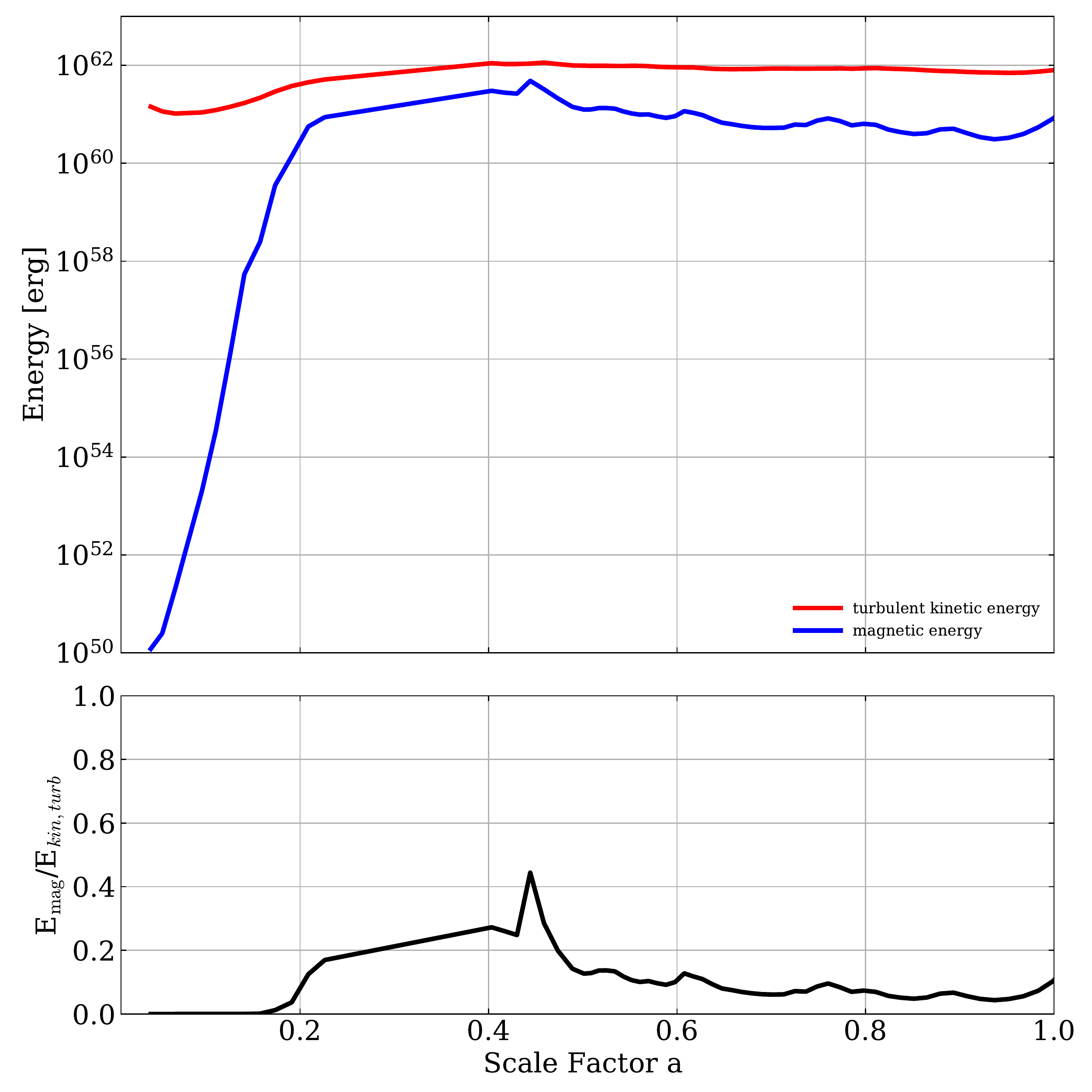}
    \caption{Time evolution of the magnetic energy (blue) and turbulent kinetic energy (red). The magnetic energy increases exponentially from the start of the simulation at redshift 300 to redshift 4 (scale factor 0.2), after which it is increasing linearly until redshift 2, after which it is decreasing towards redshift zero. The peak of magnetic energy at redshift 2 comes from intensive gravitational contraction and merger activity at these times, where the energy fraction between magnetic and turbulent energy reaches a maximum peak of around 40 per cent. Towards lower redshift, the system settles towards a fraction of around 10 per cent.}
    \label{fig:evo}
\end{figure}

We ran the MHD simulation with a rather low magnetic diffusivity ($10^{27}$ cm$^{2}$ s) and it has been shown in previous studies that a higher value for the magnetic diffusivity ($10^{28}$ cm$^{2}$ s) can decrease the magnetic field strength by a factor of 1.5 to 2 \citep[see][]{Bonafede2011, SteinwandelClusterpaper}. In this simulation, we chose a lower value for the diffusion to probe an upper limit for magnetic fields that can be expected in galaxy cluster simulations with particle-based techniques. Our future work will be centered around cosmic rays and their non-thermal emission, for which the magnetic field is an essential plasma astrophysical property, and even at low diffusivity, our method is over-shooting the magnetic field strength predicted in cluster centers based on RM signatures only by a factor of around two.  
In \citet{SteinwandelClusterpaper}, we also demonstrated evidence for a small-scale turbulent dynamo driven by ICM turbulence that is injected via merger shocks. While the process is similar to our high-resolution simulation, it is not the main aspect of this work where we are more generally interested in the general ICM properties and their plasma astrophysical context. We will investigate the detailed amplification process via power-spectra and magnetic tension in future work and instead focus on the general field structure in this paper first. 
However, for reference, we show the time evolution and the exponential build-up of magnetic energy as a function of scale factor in Fig.~\ref{fig:evo}. In blue we show the magnetic energy and in red the turbulent kinetic energy in the system. The magnetic energy grows exponentially between redshift 9 and 4 and is weaker (linear) at lower redshifts from redshift 4 to 1.5. The magnetic field energy peaks around $10^{62}$ erg. We note that the cluster undergoes very heavy merger activity from redshift 3.7 to redshift 1.3 where it assembles the majority of its mass. After redshift 1.5, the magnetic energy is slightly decreasing. This happens because the cluster evolves from a high-density accretion state to a lower-density virialized condition. We note that we find good agreement between our previous simulations with these fundamental predictions of dynamo theory but we were not able to recover the observed field strengths of systems such as the Coma cluster for which our simulations overpredict the magnetic field by a factor of 2-3. In our previous work, we extensively investigated this by changing the initial seed field, the thermal conduction prescription, and the magnetic diffusivity of our non-ideal MHD solver. While we found some indication that the initial seed field can shift the central magnetic field strength in the cluster at redshift zero, consistent with the limit of adiabatic compression the trend of the larger central magnetic field strength in comparison to Coma remains. This is also the case for our 250X-MHD simulation that is producing central magnetic field strengths that are a factor of 3-4 larger than the central field of Coma as predicted by observations \citep[see][]{Feretti1995, Bonafede2009, Bonafede2010, Bonafede2013}. However, from a theoretical perspective, the higher magnetic field strength is justified when considering the radial trends of $\beta$ and the radial trends of the energy densities. First, the radial trend of beta reveals that the thermal pressure is dominated by  roughly 1.5 orders of magnitude in the cluster center and up to 2.5 orders of magnitude in the cluster outskirts, which is characterizing our simulated ICM as typical high-$\beta$ plasma dominated by the thermal component. Furthermore, it is interesting to point out that the cluster is in rough equilibrium at redshift zero where the total kinetic energy is in equipartition with the thermal component and the magnetic pressure is in equipartition with the turbulent kinetic energy (on the smallest scales). 

\begin{figure*}
    \centering
    \includegraphics[width=0.99\textwidth]{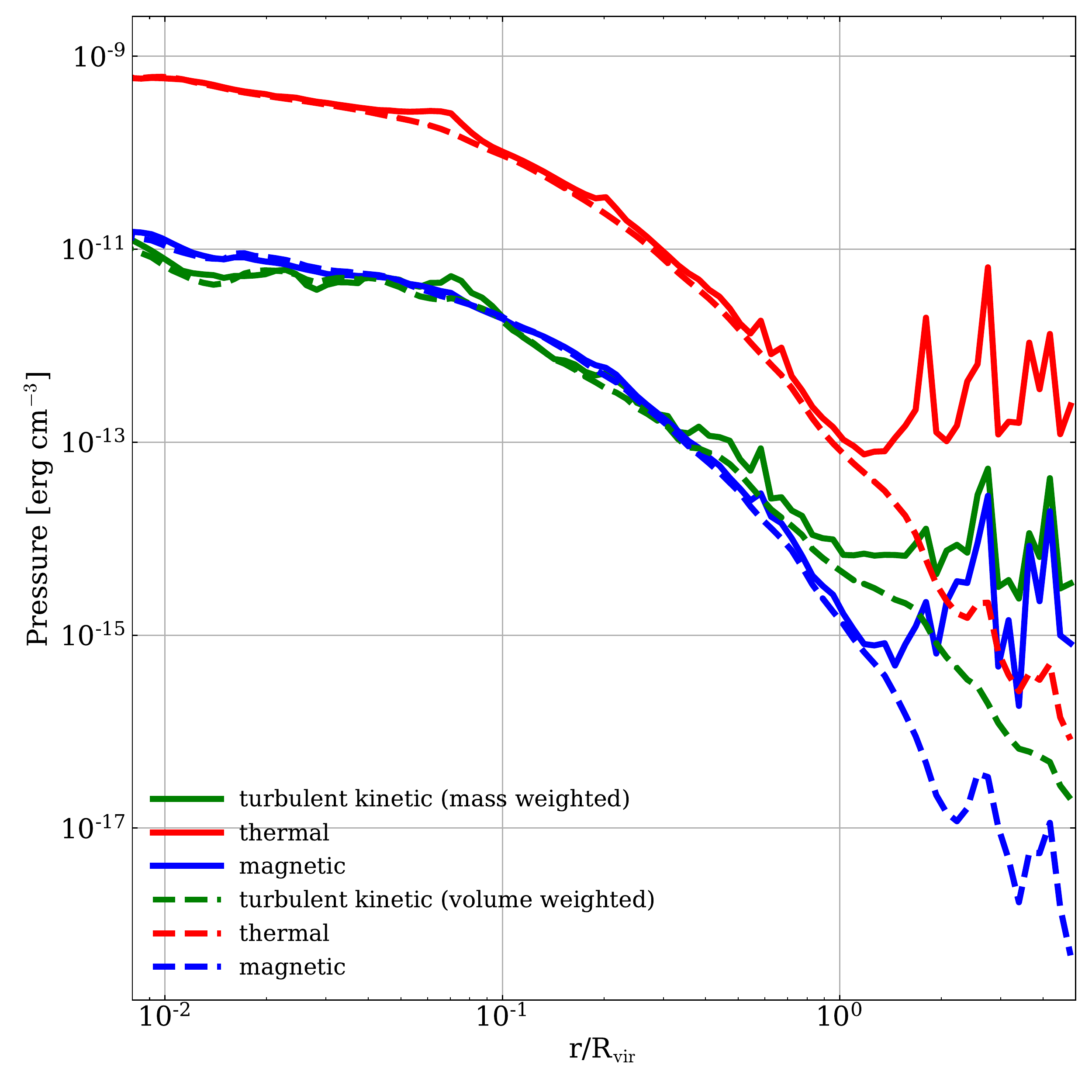}
    \caption{Radial profiles of the different pressure components that support the ICM. Red is the thermal pressure, green is the small-scale turbulent kinetic pressure, and blue is the magnetic pressure. The latter is in rough equipartition with the turbulent kinetic pressure which is in good agreement with the general predictions of dynamo theory.}
    \label{fig:pressure_pfrofiles}
\end{figure*}

In Fig.~\ref{fig:pressure_pfrofiles} we show the radial pressure profiles of the cluster at redshift zero for turbulent kinetic energy (green), magnetic pressure (blue), and thermal pressure (red). The dashed lines mark the volume-weighted quantities for completeness. On small scales, turbulent kinetic energy and magnetic energy are in equipartition. This strongly suggests that at redshift zero there is a fully saturated small-scale turbulent dynamo. The magnetic pressure and the turbulent pressure drop simultaneously towards the outskirts until 1.0 R$_\mathrm{vir}$ where both flatten (mass-weighted profiles) with a ratio of P$_\mathrm{turb}$/P$_\mathrm{B} \sim 100$. The volume-weighted profiles continue to drop. The magnetic field strength in the outskirts of the cluster is fluctuating around $5 \times 10^{-8}$ G (mass-weighted). In the mass-weighted profiles beyond 2 R$_\mathrm{vir}$, we find spikes in both turbulent kinetic pressure and magnetic pressure that are present due to the in-falling sub-structure around the most massive halo in our simulation. These are apparent only in the mass-weighted prescription and vanish by the volume weighting that naturally smooths over the finite size of halos and subhalos in the simulation. 

We note that the turbulent pressure in the center is slightly lower than the magnetic pressure by roughly a factor of 1.2 to 1.5. The reason for this lies likely in the nature of our classification for ``turbulent kinetic'' pressure which we define as $1/2 \rho v_\mathrm{rms}^2$, where $v_\mathrm{rms}$ is the random motion that remains after subtraction of the bulk motion within the kernel. That procedure is not exact and the error bar on this is at least a factor of 2. Hence, the error bar on the turbulent pressure is at least a factor of 4. Given these error bars, we are confident to make the statement that magnetic pressure and turbulent pressure are in equipartition in the cluster center. 

\begin{figure*}
    \centering
    \includegraphics[width=0.49\textwidth]{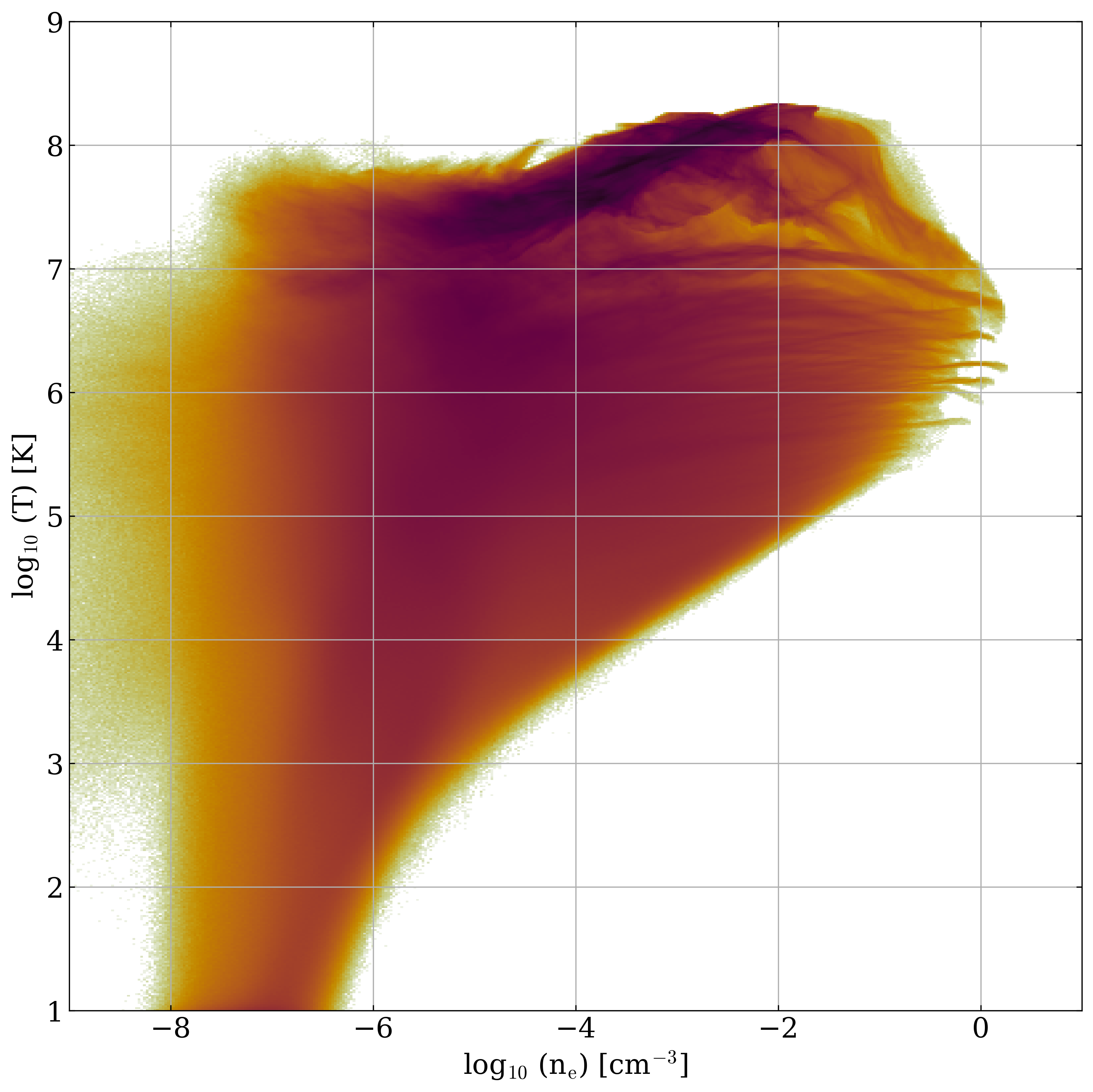}
    \includegraphics[width=0.49\textwidth]{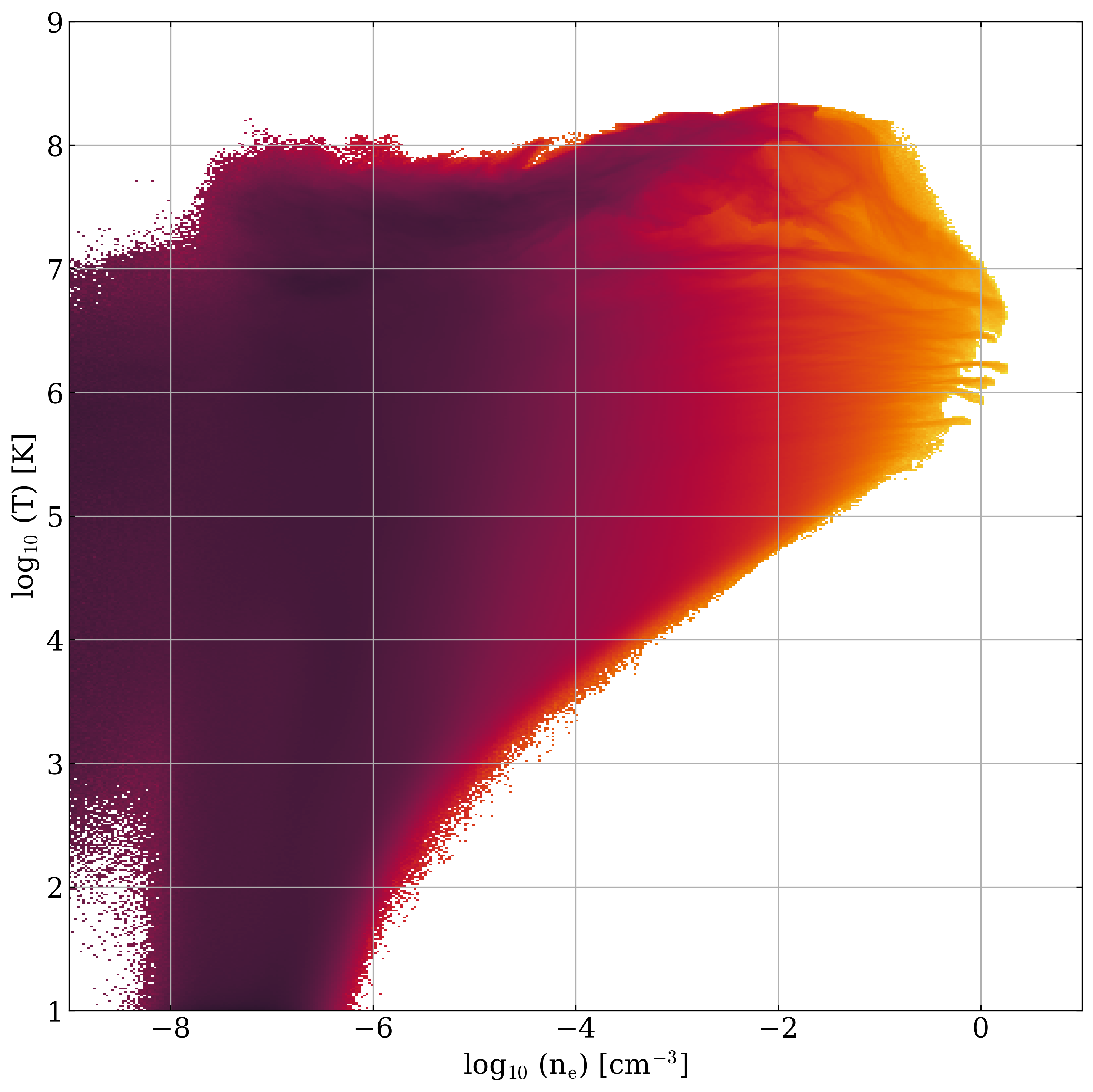}
    \caption{Mass-weighted (left) and volume-weighted (right) phase-diagrams of density and temperature. We find that most of the mass in the ICM is located in a density regime between $10^{-6}$ and $10^{-2}$ cm$^{-3}$ and temperatures between $10^7$ and $5 \times 10^8$ K. We note that the low-density part of the ICM in the outskirts of the cluster is adiabatically cooled down to very low temperatures, but the mass relative to the high-density ICM is negligible. Most of the volume of the ICM is occupied by gas with densities below $10^{-4}$ cm$^{-3}$.}
    \label{fig:phase_temp}
\end{figure*}

\begin{figure*}
    \centering
    \includegraphics[width=0.49\textwidth]{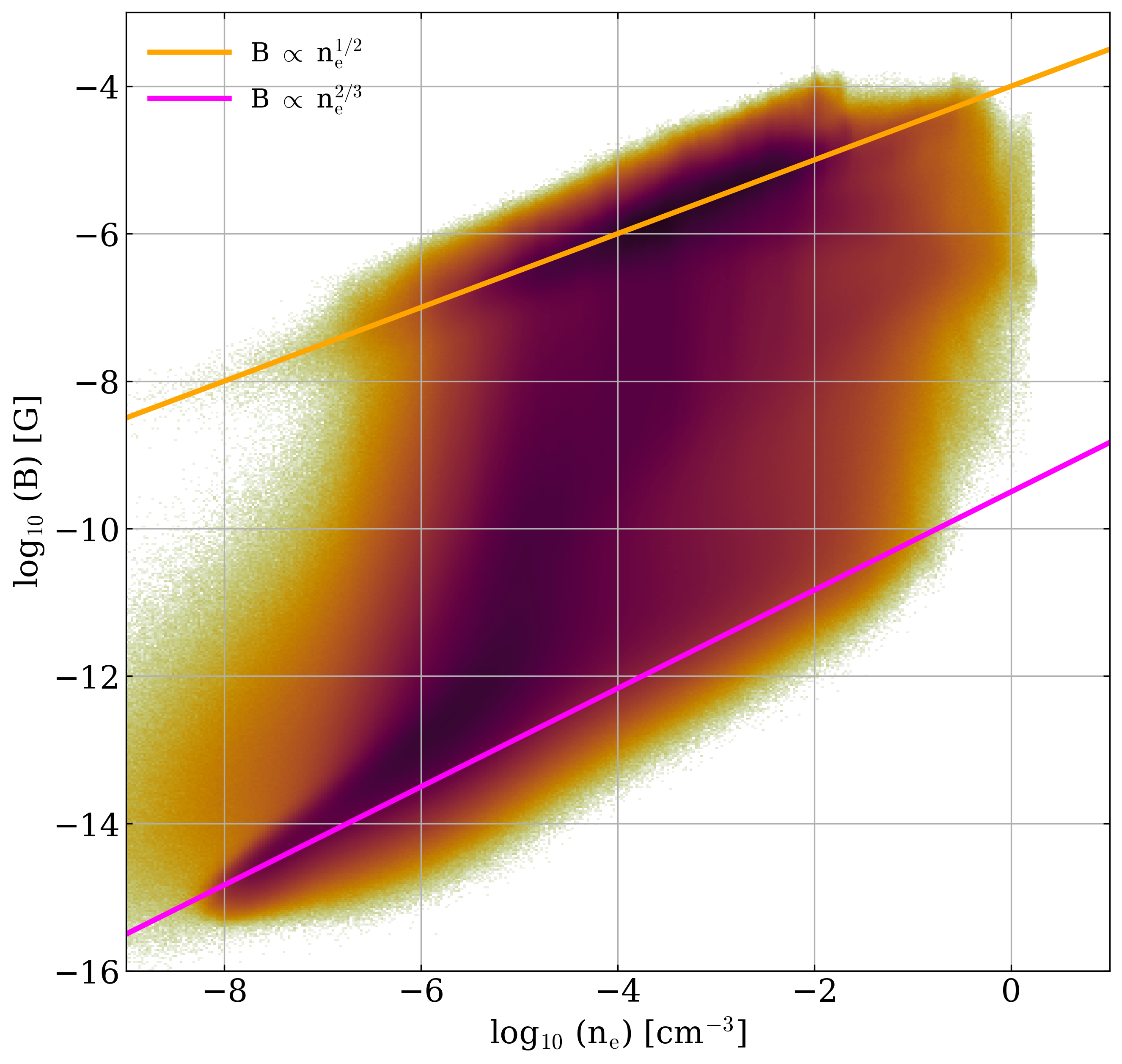}
    \includegraphics[width=0.49\textwidth]{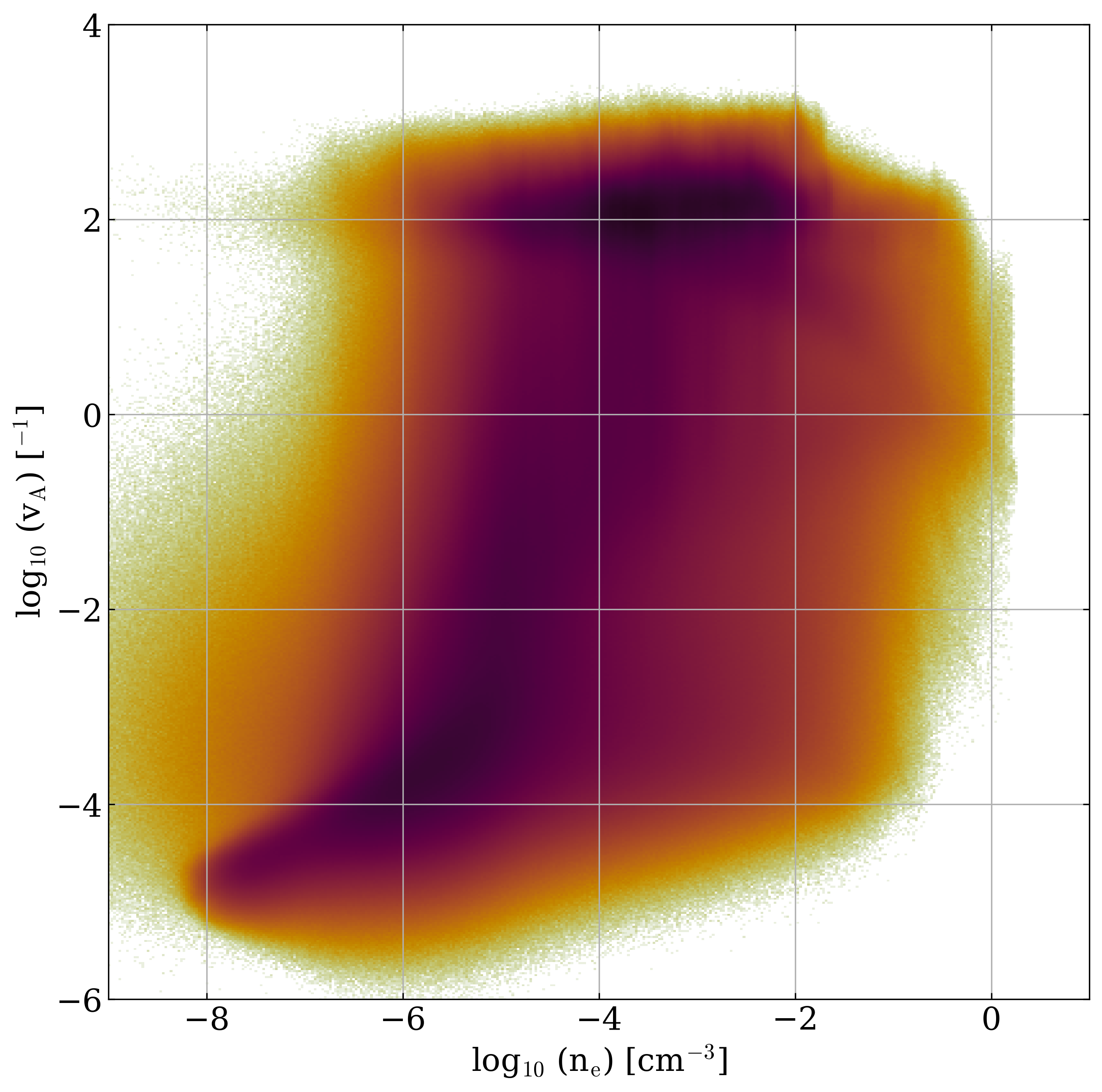}
    \includegraphics[width=0.49\textwidth]{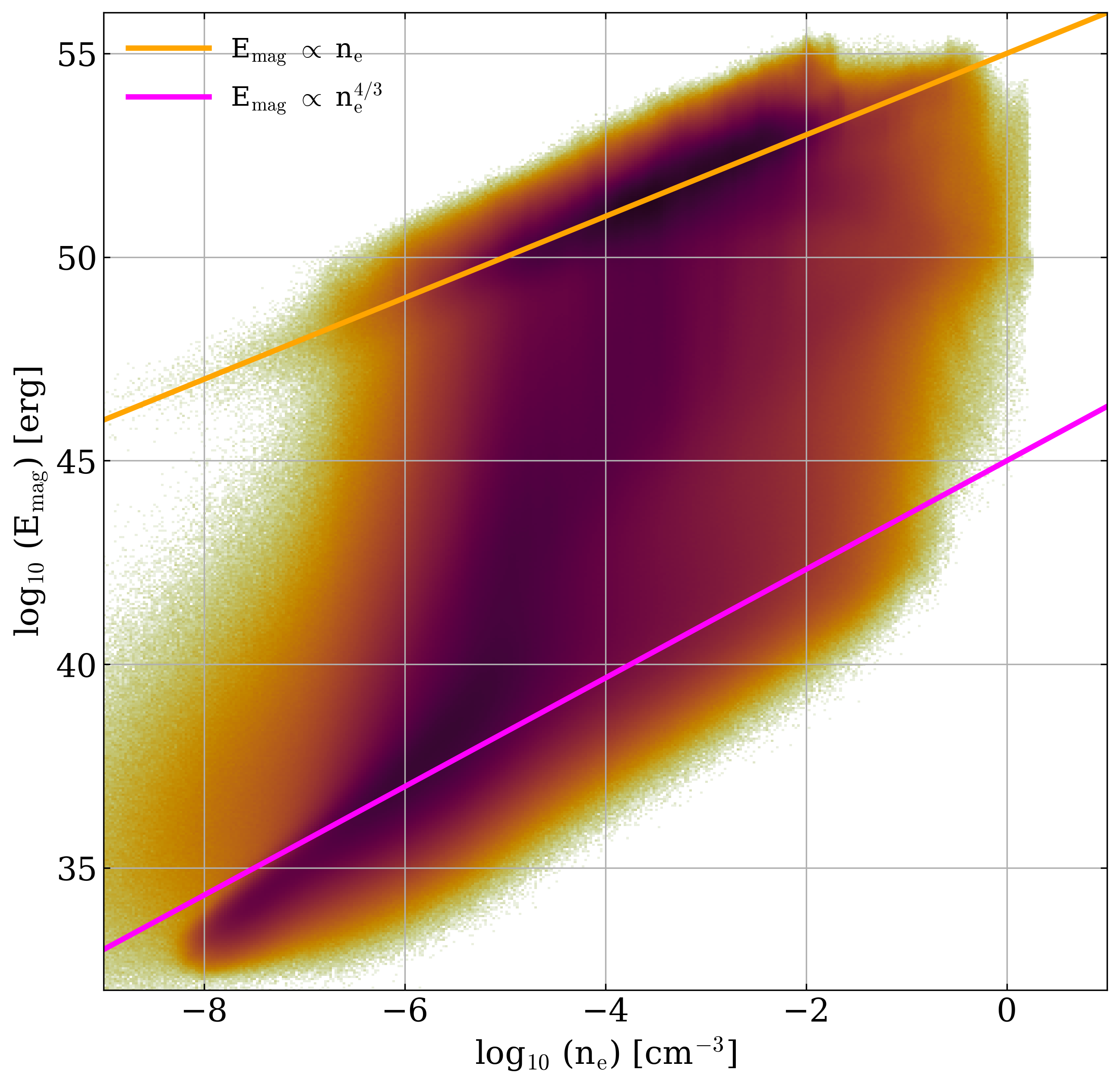}
    \includegraphics[width=0.49\textwidth]{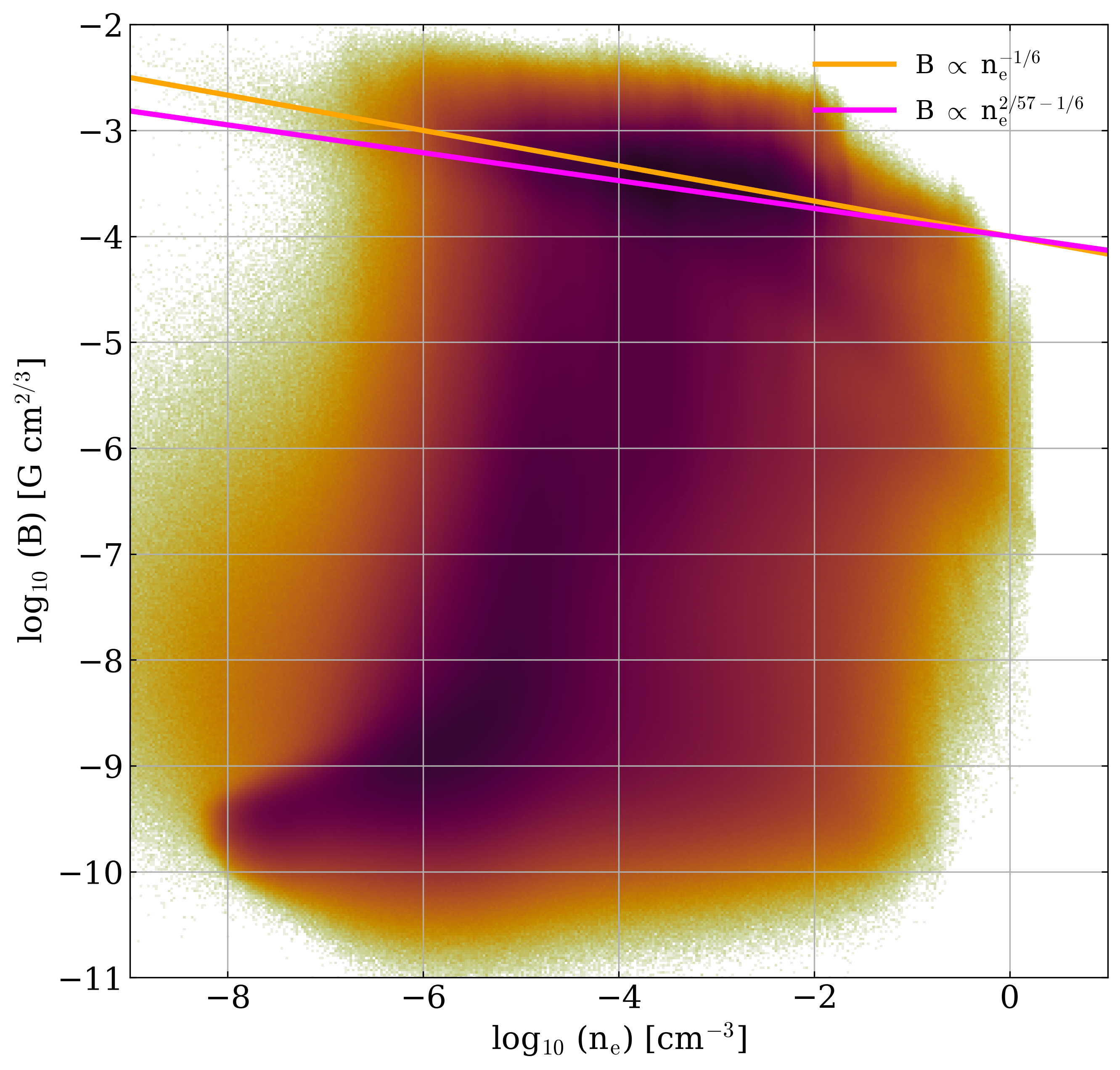}
    \caption{Joint 2d PDFs for the density-magnetic field (top left), density-magnetic energy (top right), density-magnetic field, normalized to flux freezing limit, and density-Alfv\'{e}n velocity. We find that the magnetic field and magnetic energy scale with the relations for flux freezing collapse in the weak field limit and a saturated dynamo in the strong field limit where gravitational collapse proceeds at constant Alfv\'{e}n-velocity. The PDF normalized to flux freezing limit reveals that in practice it is hard to distinguish between the saturated dynamo (-1/6 scaling) and the reconnection diffusion limit (2/57 -1/6).}
    \label{fig:2dpdfs}
\end{figure*}

In Fig.~\ref{fig:phase_temp} we show the density-temperature phase-space diagram for our simulated system at redshift zero mass (left) adn volume weighted (right). The cluster reaches characteristic temperatures of around $10^8$ K at a density of around $10^{-2}$ cm$^{-3}$. Above these densities, we observe a slight decline in temperature towards $10^7$ K, likely related to the steeper entropy profile in the cluster center. 

In Fig.~\ref{fig:2dpdfs} we show a selection of joint 2d PDFs of several central quantities that are correlated with the magnetic field as a function of electron number density. All these PDFs are computed at redshift zero with all $\sim 10^{9}$ gas particles in the simulation. The top left panel shows the magnetic field strength as a function of electron number density. The magenta line shows the adiabatic (flux freezing) compression regime where $\mathbf{B}$ scales as n$_{e}^{2/3}$. The golden line shows what is typically referred to as the saturated dynamo regime where $\mathbf{B}$ scales as n$_{e}^{1/2}$. Generally, we find that the system follows the adiabatic compression limit in low-density and weakly magnetized gas. At higher magnetic field strengths (and densities) we find that the system follows the scaling of the saturated dynamo. In the upper left panel of Fig.~\ref{fig:2dpdfs} we show that this saturated dynamo regime is equivalent to gravitational collapse at constant Alfv\'{e}n-velocity, which is a distinct feature for strong magnetic fields over at least four orders of magnitude in density (from 10$^{-6}$ to 10$^{-2}$ cm$^{-3}$). We explicitly note this here because it is often overlooked. Since Alfv\'{e}n-speed scales as $\rho^{-1/2}$ and we obtain a saturated dynamo regime that scales as $\rho^{-1/2}$ as well, we must achieve collapse at constant Alfv\'{e}n-velocity of the densest regions of the ICM. To that end, it is worth noting that the nature of turbulence is subsonic and sub-alfv\'{e}nic. 
In the left panel of the bottom row of Fig.~\ref{fig:2dpdfs} we additionally show the 2d PDF of magnetic energy alongside the expected scalings for magnetic energy with power-law indices of 0.25 for the saturated dynamo regime and 4/9 for the adiabatic limit for completeness. Finally, we show the 2d PDF for the quantity B/n$^{2/3}$ and compare to the slopes of the saturated dynamo following \citet{Kraichnan1967, Kazantsev1968} (golden line) as well as the reconnection diffusion limit derived by \citet{Xu2020} (purple line). We note that in practice it is very hard to distinguish between these two scenarios based on the slope in this diagram alone and one would have to carry out a detailed study of the reconnection rates, which is beyond the scope of this work.

\subsection{Magnetic field structure}

\begin{figure*}
    \centering
    \includegraphics[width=0.99\textwidth]{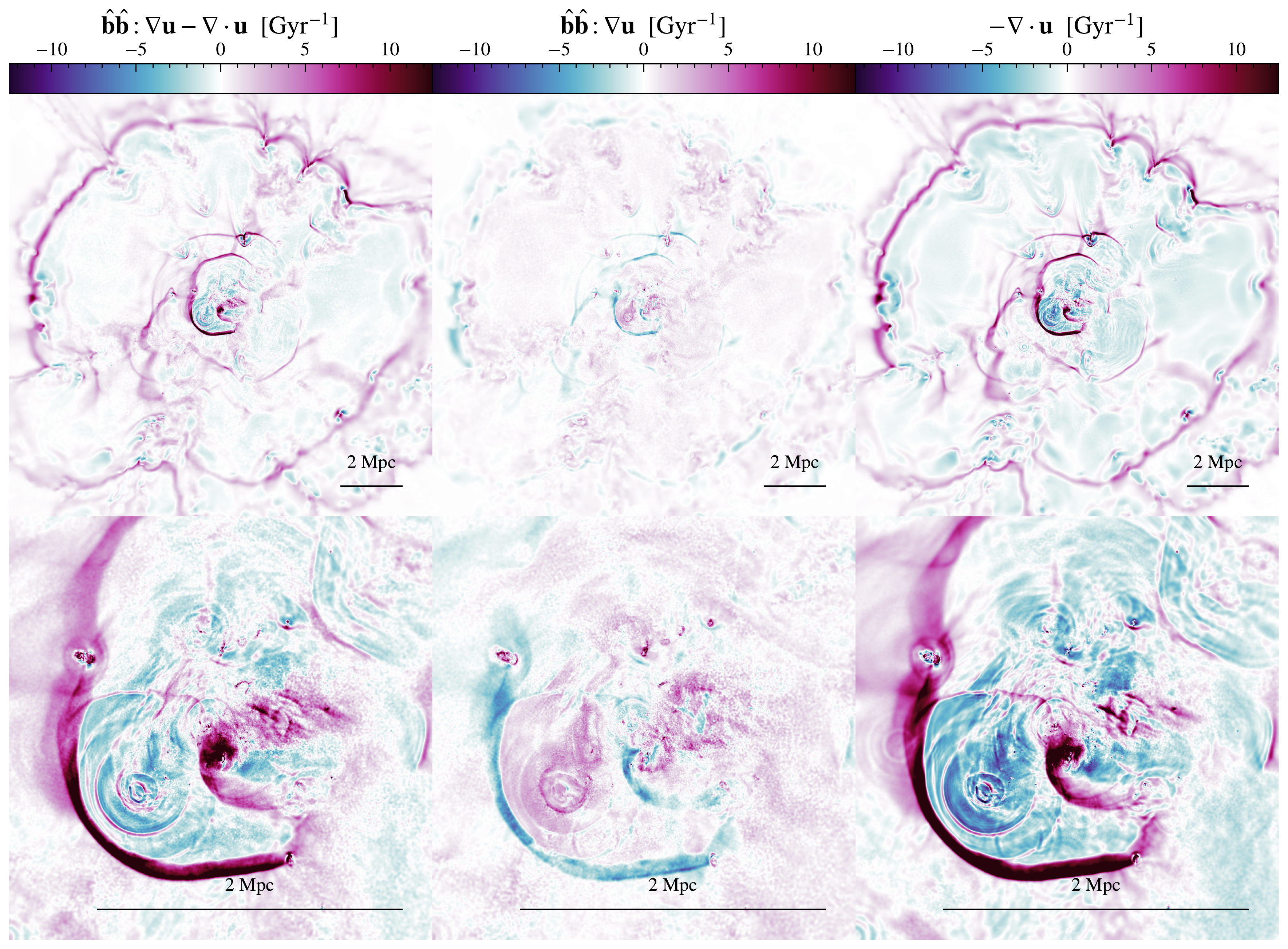}
    \caption{Total rate of change of the magnetic field (left), shearing/turbulent rate of change of the magnetic field (center), and compressive rate of change of the magnetic field (right). The top row shows the whole simulation domain, while the bottom panel is focusing on the field structure around a cold front that forms right at redshift zero through a sub-structure that is penetrating the cluster center.}
    \label{fig:bb}
\end{figure*}

In Fig.~\ref{fig:bb} we show the total rate of change of the magnetic field in the whole simulation domain (top left), as well as the rate of change, split into shearing/turbulent motions (top center) and compressive modes (top right). In the bottom row of Fig.~\ref{fig:bb} we zoom in on a cold front that forms at low redshift in the very center of the cluster when a sub-structure that is penetrating the center is compressing gas in bow-shock that moves away from the cluster center. The compressive part on the right is easier to interpret as it is simply the velocity divergence. The center panels that mark shear are harder to interpret, as we have to carry out a contraction (Frobenius norm) over the two involved second-order tensors. Hence, $\hat{\mathbf{b}} \hat{\mathbf{b}} : \nabla \mathbf{u}$ is defined as:
\begin{align}
    \hat{\mathbf{b}} \hat{\mathbf{b}} : \nabla \mathbf{u} = \sum_{i}^{3} \sum_{j}^{3} {b_{i}}{b_{j}} \partial_{i} u_{j}. 
\end{align}
where $b_i$ and $b_j$ are the components of $\hat{\mathbf{b}}$ which is itself the (normalized) three-dimensional magnetic field strength. Next, we need to understand where these terms are coming from. For this, let us consider the induction equation in the form: 
\begin{align}
    \frac{\partial \mathbf{B}}{\partial t} = \nabla \times (\mathbf{u} \times \mathbf{B}) + \eta \nabla^2 \mathbf{B}.
\end{align}
and drop the second (resistive) term. If we dot this equation with $\mathbf{\hat{b}}/B$ and evaluate the first term on the right-hand side using common vector identities as well as using the definition of the Lagrangian derivative 
\begin{align}
    d/dt = \partial/\partial t + \mathbf{u} \cdot \nabla 
\end{align}
we will get the following form of the induction equation:
\begin{align}
    \frac{1}{B} \frac{d\mathbf{B}}{dt} = \hat{\mathbf{b}} \hat{\mathbf{b}} : \nabla \mathbf{u} - \nabla \cdot \mathbf{u}.
    \label{eq:bb}
\end{align}

The first term is essentially measuring the shear (induced by the presence of the magnetic field), and the second term is the velocity divergence that indicates adiabatic compression and decompression respectively, depending on the sign. In summary, this allows us to study regions in the ICM where magnetic field growth (or suppression for that matter) is driven by shear/turbulence (first term in eq.~\ref{eq:bb} ) or the compressibility of the gas (second term in eq.~\ref{eq:bb}). Hereby, it is important to note that the second term in eq.~\ref{eq:bb} is typically dropped as many ICM studies are carried out under the assumption of incompressibility \citep[e.g][]{Squire2023} that yields $\nabla \cdot \mathbf{u} \approx 0$. Our simulation results from Fig.~\ref{fig:bb} indicate that in a volume-filling interpretation, this is actually a very good assumption. The only regimes in which this is violated are at the shock fronts in the ICM due to merger activity which is traced excellently by $\nabla \cdot \mathbf{u}$. We find that the magnetic field is strongly increasing at the shock fronts due to compression. However, the bulk of the amplification in the vast majority of the volume is driven by shear as indicated by the top center panel of Fig.~\ref{fig:bb}. We note that the total rate of change in the volume is generally speaking low at redshift zero and only the shocks are able to produce a positive rate of change of the magnetic field strength that is significant. However, these features are obviously highly transient and their magnetic energy will be dissipated after the shocks dissolve. 

\begin{figure*}
    \centering
    \includegraphics[width=0.49\textwidth]{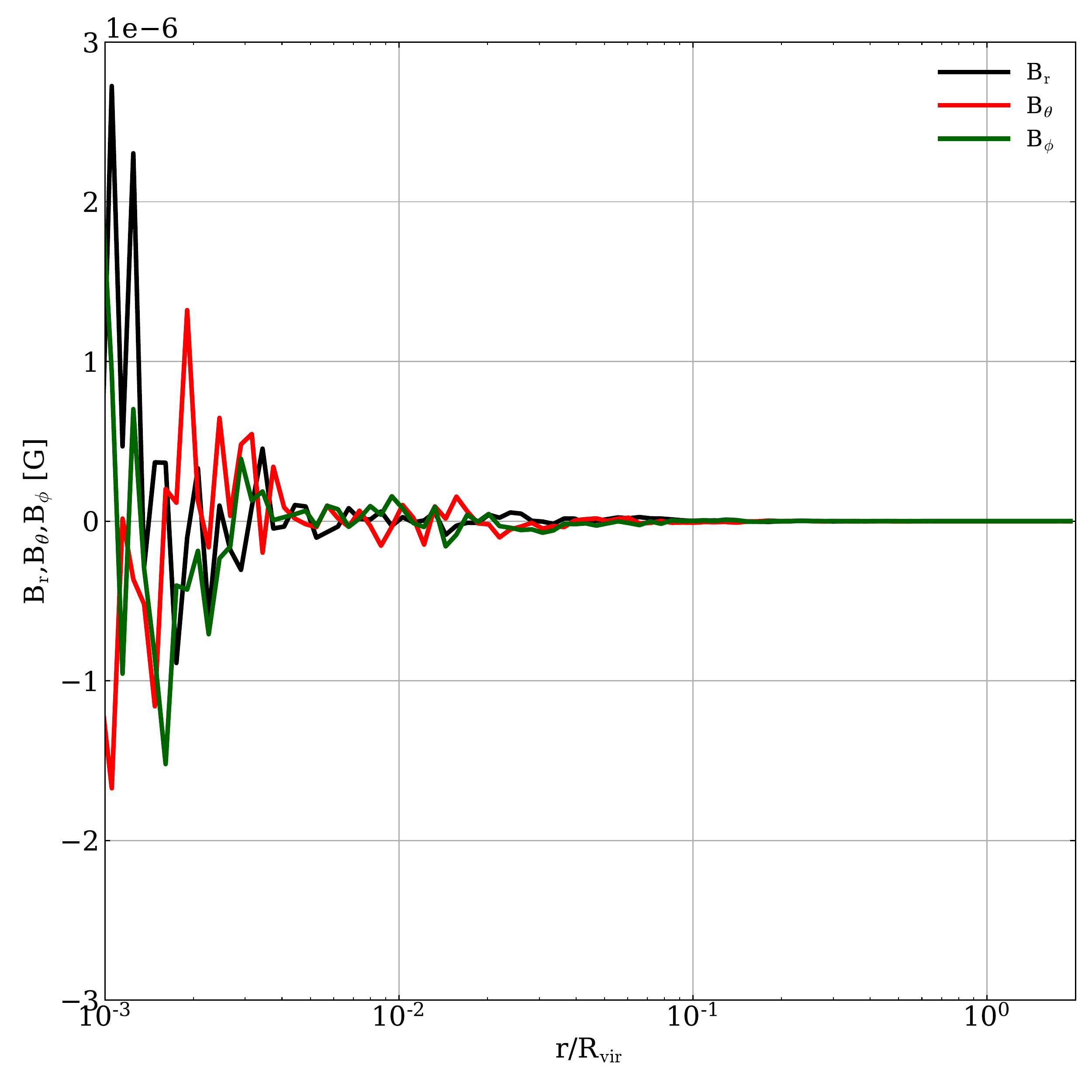}
    \includegraphics[width=0.49\textwidth]{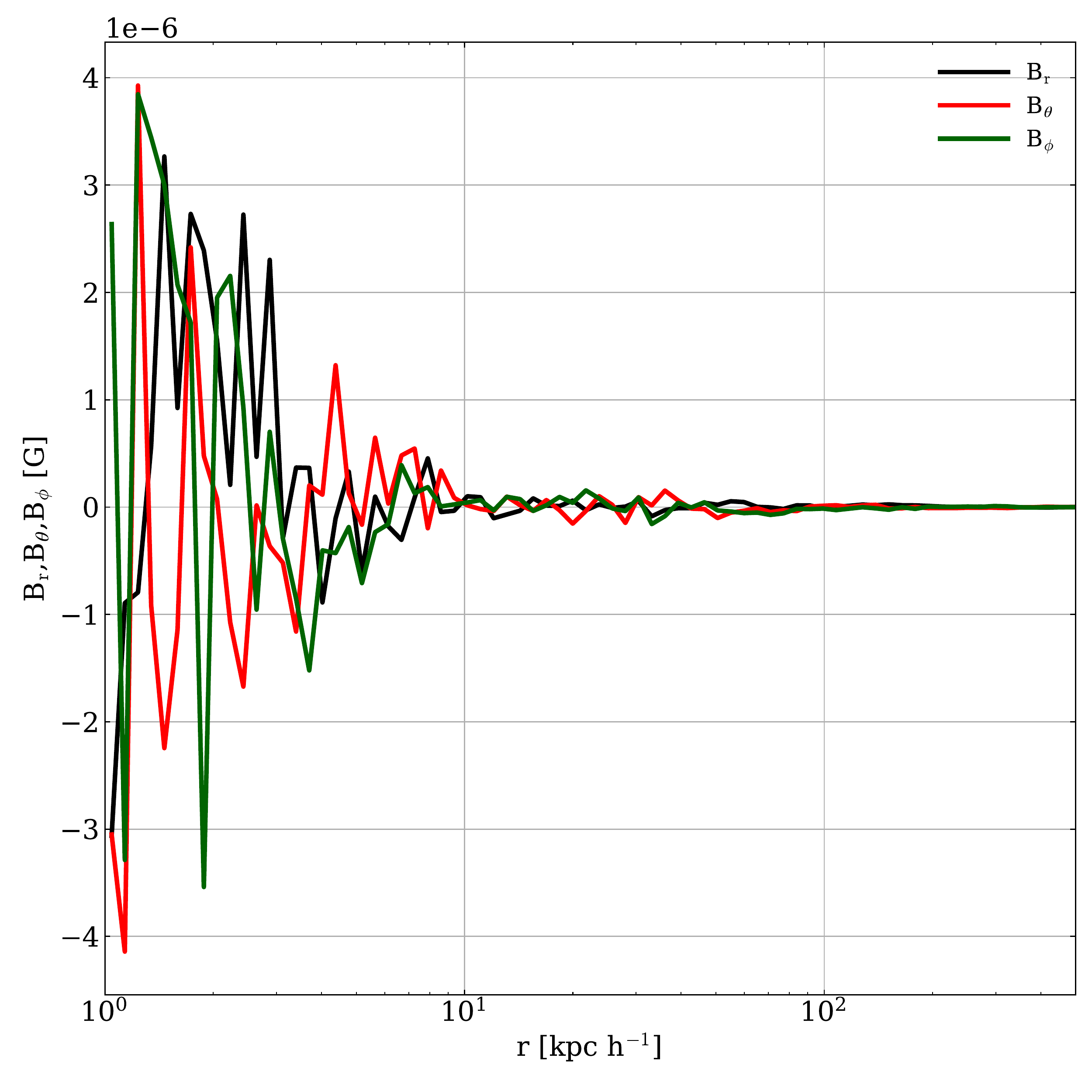}
    \caption{Left: Radial profiles of the radial, toroidal, and azimuthal components of the field out to the virial radius at 2300 kpc. Right: Zoom-in of the radial profiles on the innermost 500 kpc of the simulated cluster. At large radii, the components average out, indicating a high symmetry in these parts of the cluster. In the central parts, the radial and toroidal components dominate the azimuthal component, but the trend is weak, indicating that the field structure is largely set by large-scale random motion induced but the structure formation process.}
    \label{fig:b_structure}
\end{figure*}

In Fig.~\ref{fig:b_structure} we show the radial, and angular components of the magnetic field out to the 2 R$_\mathrm{vir}$ (left) and the innermost 500 kpc of the cluster (right). Generally, we find that the radial component and the angular components show field reversals on the scale of $5-10$ kpc in the central region and approach zero with increasing distance from the center. This simply means that in the cluster outskirts, the positive and negative orientations of each component average out because they are isotropized by large-scale bulk in and/or outflow towards (away) from the cluster center. However, the substructure in-falling onto the central regions of the cluster drives strong merger shocks towards the outskirts which breaks this symmetry and injects turbulence that ultimately drives a small-scale turbulent dynamo that converts radial magnetic field to angular field. This is apparent due to the fact that radial and angular components show some evidence for alternating field reversals. That the field is organized on the larger scales by the structure formation process becomes apparent when we consider the 1D differential PDF of the different components of the magnetic field that is  generally symmetric. However, we do find a slight excess radial field at around 10$^{-6}$ compared to the angular components of the magnetic field.

In Fig.~\ref{fig:b_rotb} we show the absolute value of $\mathbf{J} = \nabla \times \mathbf{B}$ in the left panel and its z component $J_{z} = \partial_\mathrm{x} B_\mathrm{y} - \partial_\mathrm{y} B_\mathrm{x}$ to further quantify the turbulent small-scale structure of the magnetic field. Specifically, J$_{z}$ indicates some evidence of magnetic reconnection in the ICM. To which degree this is resolved in this simulation will be investigated in greater detail in future work. However, the small-scale structure we find indicates a structure known from ``Plasmoids'', although we note that they are likely unresolved here.

\section{Discussion}

\begin{figure*}
    \centering
    \includegraphics[width=0.49\textwidth]{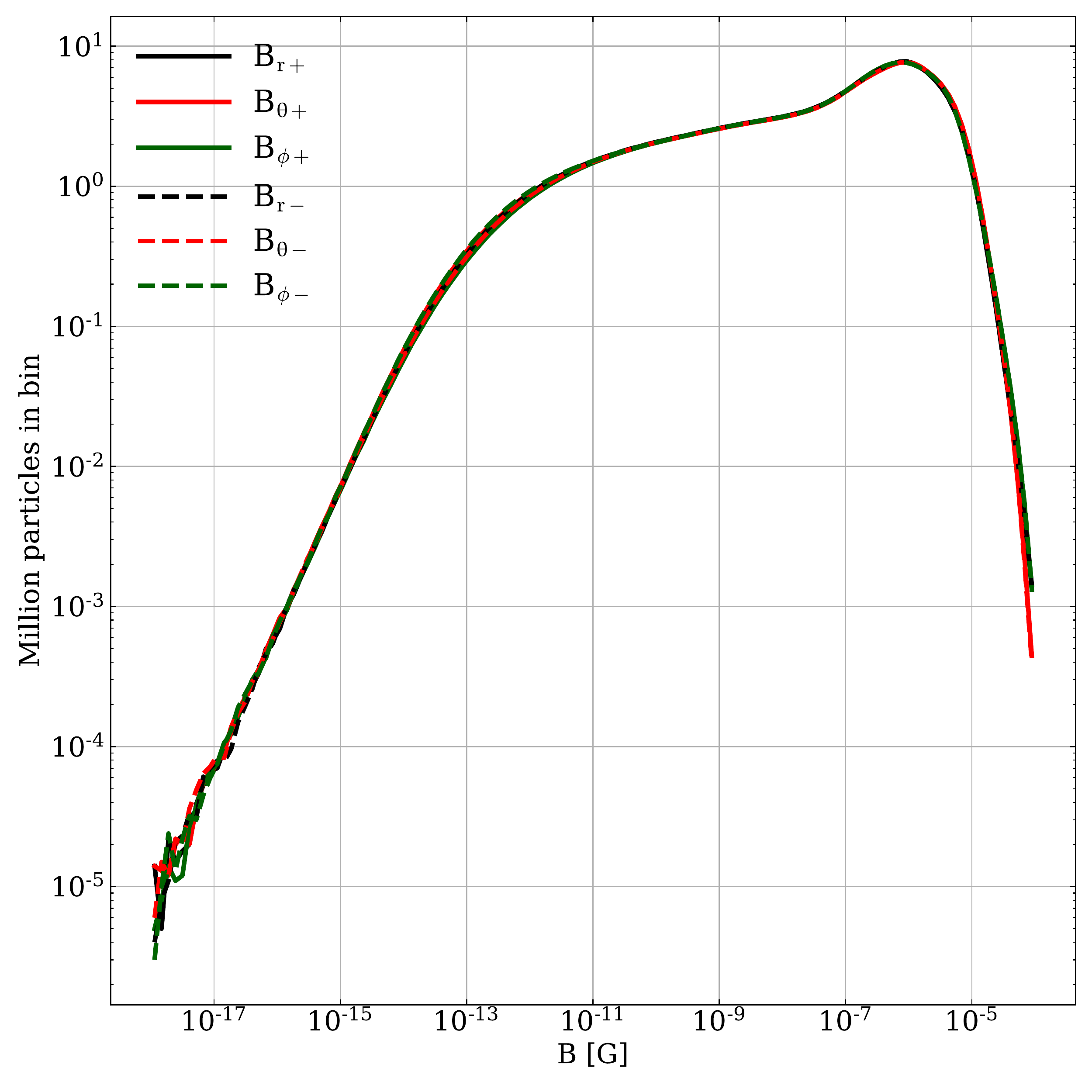}
    \includegraphics[width=0.49\textwidth]{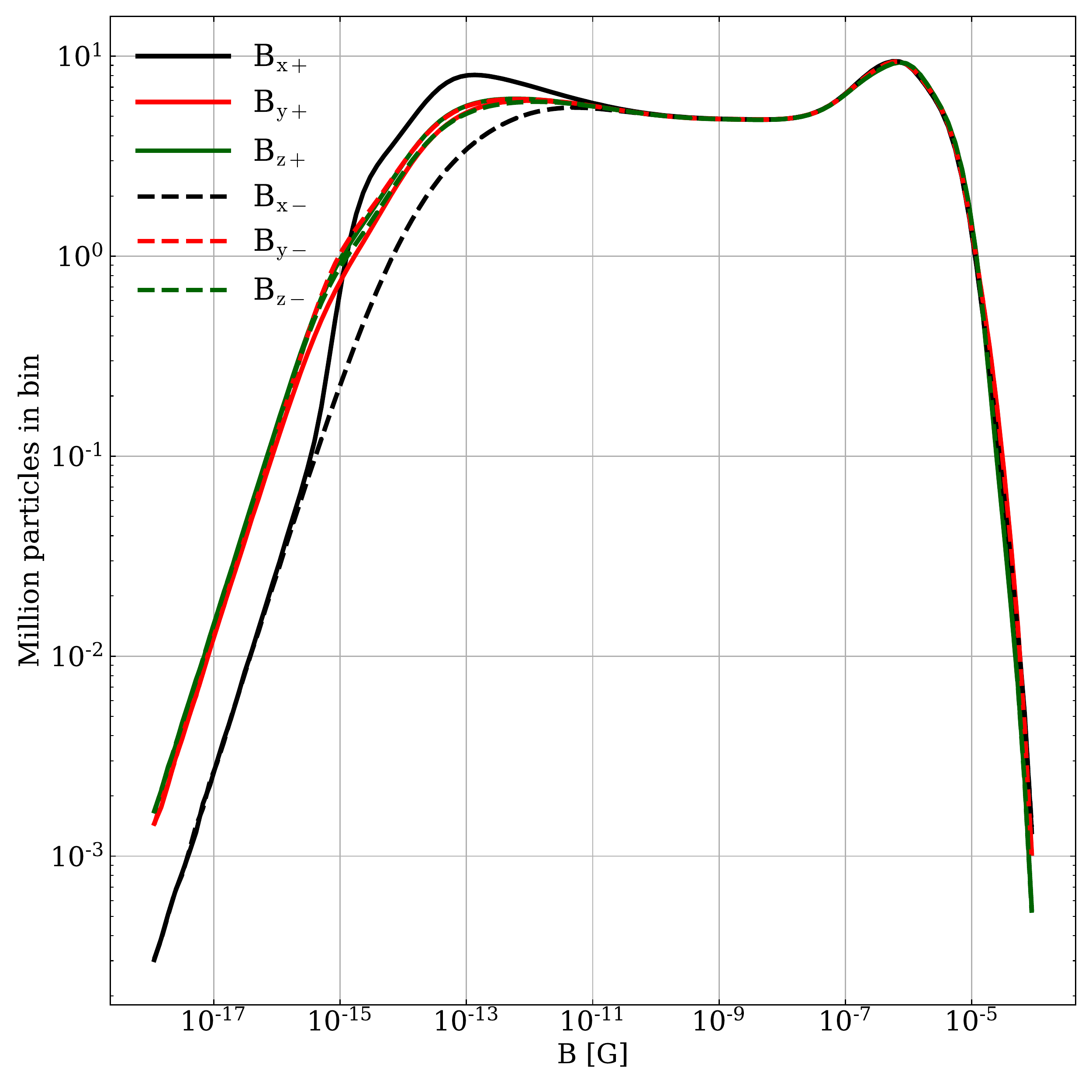}
    \caption{Left: One dimensional PDF of the different field components B$_\mathrm{r}$, B$_\mathrm{\theta}$ and B$_{\phi}$. These indicate that the field is symmetric in all its components, which means that the field is ordered by large-scale random motions, and the final field structure is decoupled from the field structure that is enforced in the early formation process by the dynamo process. Right: Same PDFs for the B$_\mathrm{x}$, B$_\mathrm{y}$, B$_\mathrm{z}$ which indicates that the field in the ``Void'' regions retains some memory of its initial configuration as there is a bias towards a positive B$_\mathrm{x}$ component which was the original orientation of the seed field. However, this is only true for regions that are dominated by ``comoving'' expansion.}
    \label{fig:b_1dpdfs}
\end{figure*}

In this section, we will discuss our results and compare them to other relevant work centered around numerical simulations of magnetic fields in galaxy clusters. We will split this section into three parts, discussing the general morphological features of the simulated ICM, the amplification of the magnetic field, and its saturation at a fraction of turbulent kinetic energy. Finally, we will discuss the structure and the development of the small and large-scale magnetic field. 

\subsection{Consequences of resolving the mean free path of the electrons \label{sec:mean_free_path}}

One important features of the presented simulation is the fact that it operates in a resolution regime where the resolution for most of the Lagrangian mass is of the order of, or better than the electron mean free path. We show this very important characteristic in Fig.~\ref{fig:mean_free_path}. Hereby, it is important to note that this will obviously not be the case everywhere in the simulation and we restrict this statement for the typical ICM densities and temperatures between $10^{-7}$ cm$^{-3}$ and $10^{-2}$ cm$^{-3}$, as well as $10^{6}$ K and $10^{8}$ K. Most of the mass of the simulation is located in this regime and thus it is appropriate to center the discussion around that regime. The advantage of a Lagrangian code, such as ours is that we have very good resolution in dense regions where we find that the mean free path is generally much smaller than the resolved resolution scale. The mean free path will be large if either temperature is high or density is low and thus we find that the mean free path is large in the outskirts of the cluster where we find that despite the fact that the resolution is of order 100 to 500 kpc it is still at least 2 orders of magnitude higher than the actual electron mean-free path and hence the cluster outskirt is clearly a region where the pure MHD treatment might not be a good approximation anymore and even ``kinetic aware'' theories such as Braginskii-MHD \citep[][]{Braginskii1965} might not be the appropriate limit anymore to treat that regime and fully kinetic limit would be favorable. As of now, it is questionable if this can be achieved. We note, that there is a tail with a low electron mean free path in the outskirts which sits at the edge of the zoom region and is cooled adiabatically as it expands in the Hubble flow (top right of the left panel of Fig.~\ref{fig:mean_free_path}). 

The interesting region for us in this work is located at a mean free path between 1 and a few 100 kpc (right panel of Fig.~\ref{fig:mean_free_path} where the resolution is on average at least half a dex smaller than the electron mean free path. This means that our simulation is pushing the edge for MHD in that regime and cluster simulations at our target resolution will be the ideal playground for the effect of ``kinetic aware theories''. If we were to push the resolution even further, we believe it might be necessary to adopt an MHD-kinetic hybrid approach. The question, however, is, if the effects will be strong enough to make a difference as it has been recently pointed out by \citet{Squire2023} that MHD might be a good approximation for the ICM after all. Regardless, it will be interesting to investigate the rates of strain and pressure anisotropies in future simulations of galaxy cluster formation to ``bridge the gap'' to the detailed work on plasma instabilities that can occur in the ICM.    

\subsection{Morphology and structure of the ICM} 

We extensively probed the structure and morphology of our simulated ICM with a specific focus on the simulation output at redshift zero and the present turbulent field structure. We present central ICM properties of the cluster in Fig.~\ref{fig:cluster} and Fig.~\ref{fig:cluster_yz} where we show the density and temperature field of the simulated system, in the left and middle panel of Fig.~\ref{fig:cluster}. At redshift zero the cluster is representing a relaxed gravitational system with a massive central halo with a mass of $\sim 2 \times 10^{15}$ M$_{\odot}$ that is surrounded by a number of massive group-sized objects with masses of up to $\sim 5 \times 10^{13}$ M$_{\odot}$ and one smaller galaxy cluster with a mass of $\sim 1.1 \times 10^{14}$ M$_{\odot}$. The morphological structure within the virial radius of the most massive halo ($\sim 3.1$ Mpc) is governed by a number of internal shocks that is arising from the in-falling substructure into the most massive halo, which appears to be the major source of turbulence in the ICM in the absence of the feedback of AGN. This results in a turbulent structure within the virial radius and a smoother distribution beyond, dominated by the individual peaks of the in-falling sub-structure. The turbulent small-scale structure is very apparent in Fig.~\ref{fig:cluster_yz} where we show a very thin slice of 100 kpc height along the x-direction that reveals strongly magnetized filamentary structures in the innermost Mpc of our simulated cluster. Despite the fact that these are smaller-scale with respect to the virial radius of the system, they still appear to be correlated over a length scale of at least a few 100 kpc where these regions reach field strengths of up to 20 $\mu$G. This is in agreement with the radial profiles of the magnetic field strength of our simulated system in the upper left panel of Fig.~\ref{fig:radial_profiles} that indicates that the magnetic field strength is reaching 20 $\mu$G in the center of the cluster and dropping to a field strength of around 4 $\mu$G at a scale of around 0.2 R$_\mathrm{vir}$ which corresponds to a physical scale of around 600 kpc. At larger scales, the field drops rapidly to a field strength of around 0.03 $\mu$G at the virial radius which corresponds to a physical scale of 3100 kpc. With respect to our own earlier work this is encouraging since the radial profiles seems to be converged in comparison to our lower resolution runs that we put forward in \citet{SteinwandelClusterpaper}. However, the discrepancy with ICM magnetic field models, that are put forward with grid codes such as \textsc{Enzo} remain and our simulations over-predict the observed magnetic field in galaxy clusters by roughly a factor of three compared to the observations in Coma \citep[e.g.][]{Bonafede2011, Bonafede2013}. Moreover, we also can explain this based on the simulation data at hand. For instance when we consider the radial profiles of temperature and entropy in the bottom panels of Fig.~\ref{fig:radial_profiles} we find a steep drop in the temperature and entropy in the center of the cluster. This becomes even more apparent if we consider the 2d PDF of density and temperature in Fig.~\ref{fig:phase_temp}. The drop in temperature and entropy leads to an increase in density towards the center as well, a well know problem for non-radiative simulations of galaxy clusters with SPH-methods. The effect is actually very clearly illustrated by the resolution study in our previous paper as presented in Fig.~4 of \citet{SteinwandelClusterpaper}. Hence, it is easy to show that the increase of the field in the central part of our simulated cluster comes from the increase in density due to a decreasing entropy and temperature profile. The cluster magnetic field then just follows the increase of the density field in the adiabatic compression limit. Finally, we want to highlight that given these limitations of the simulation, we still find good agreement of our simulated ICM with a high-$\beta$ Plasma with a central value of around 50 that reaches a few hundred a the scale of the virial radius and drops to around 1000 beyond the virial radius. These values are not all unrealistic when compared to the previous assumptions of the ICM as a high beta plasma. 

\subsection{Dynamo amplification of the 
magnetic field}

We will briefly discuss the dynamo amplification of the field in the ICM but will postpone a more detailed spectral analysis via power spectra to future studies when we have a larger set of simulations available at our target resolution. Similar to our lower resolution runs that we presented in \citet{SteinwandelClusterpaper} we find that the magnetic field is increasing rapidly from the starting redshift 310 to around redshift 4 where we find a peak of the magnetic energy as shown in Fig.~\ref{fig:evo}. It is interesting to point out that the total magnetic energy is always lower than the turbulent kinetic energy and at redshift zero we find a saturation value that of around 10 per cent. This is in good agreement in comparison to earlier work on dynamo theory. For instance \citet{Schober2015} investigated the saturation level of the turbulent dynamo for different magnetic Prandtl-number (Pm) regimes for $Pm \ll 1$ and $Pm \gg 1$ for different assumptions of the spectrum of the underlying MHD turbulence and find values between 0.1 and 3 per cent for $Pm \ll 1$ and 1 and 30 per cent for $Pm \gg 1$. Our simulation is in the regime of the latter and we find a operates at magnetic Reynolds number of a few 100 to a few 1000 putting us well above the regime for resolved dynamo action under the assumption of incompressible Kolmogorov turbulence, which is a fair assumption for the ICM. Thus we could expect a saturation value for the dynamo somewhere up to a few 10 per cent. Hence the 10 per cent at redshift zero is in rather good agreement with the precision of \citet{Schober2015} given the uncertainties in the exact nature of the turbulence. We note that the saturation value seems to be larger at higher redshift reaching a peak value of around 42 per cent at redshift 1.5. \\

It is additionally important to point out that the magnetic field is in rough equipartition with the turbulent kinetic energy within the virial radius of the system, indicating one fundamental prediction of dynamo theory as shown by Fig.~\ref{fig:pressure_pfrofiles}. A more detailed study of the magnetic field and related proprieties reveals a more complete picture of the magnetization of the ICM. In Fig.~\ref{fig:2dpdfs} we show a number of these properties in 2D mass-weighted PDFs as a function of electron number density. First, the top left panel shows the magnetic field strength itself. It is important to discuss the
structure of the magnetic field strength itself. One can split the ICM into two regimes: a high-density part that is also strongly magnetized as well as a low-density part that is weakly magnetized. The dense gas is following a scaling of $B \propto \rho^{1/2}$ which is for instance predicted by reconnection diffusion in the saturated regime of the turbulent dynamo. Hence the dense regions of the ICM undergo collapse in a saturated dynamo regime. More importantly, it is interesting to point out that this directly implies collapse at constant Alfv\'{e}n-velocity which we show in the top right panel of Fig.~\ref{fig:2dpdfs}. This is not really surprising given the definition of Alfv\'{e}n-velocity as 
\begin{align}
    v_{A} = \frac{|\mathbf{B}|}{ \sqrt{4 \pi \rho}}.
\end{align}
Hence if the magnetic field is proportional to the square root of density, Alfv\'{e}n-velocity in collapsing regions needs to be constant. Essentially, this is directly enforced by the equipartition condition where we assume:
\begin{align}
    B_\mathrm{rms}^2 \propto \rho v_\mathrm{rms}^2.
\end{align}
This enforces B$_\mathrm{rms} \propto \rho^{1/2}$ and collapses at constant Alfv\'{e}n-velocity if the turbulent velocity is constant as collapse proceeds. The latter is actually nontrivial since collapse to very dense regions would transit from sub-to supersonic regimes under strong cooling. This is not an issue under fully ionized ICM conditions and turbulent velocity is roughly constant down to the highest ICM densities that we resolve. However, it has some important consequences regarding the nature of the underlying turbulence that is subsonic with a sonic Mach number of $\mathcal{M}_\mathrm{s} \approx 0.05 - 0.1$ and Alfv\'{e}n-Mach number $\mathcal{M_\mathrm{A}} \approx 0.5 - 1.5$ in the high-density part of the ICM, that is in the weakly sub- to the trans-alfv\'{e}nic regime. Additionally, turbulence in our case is strongly intermittent due to the nature of its origin by merger shocks, which makes it generally harder to identify the nature of the turbulence over $\mathcal{M}_\mathrm{s}$ and $\mathcal{M}_\mathrm{A}$. Hence, we only make a statement on these numbers at the relaxation and saturation of the system. Finally, it is important to point out that the Alfv\'{e}n-velocity, in general, is very high with values of around 200-300 km s$^{-1}$. The origin of this is the generally low densities that govern the ICM, while at the same time, the mean magnetic field remains at $\mu$G level similar to the field observed in local spiral galaxies \citep[see][for a review on this topic]{Beck2015}. It is worth mentioning that the magnetic energy is amplified by around 20 orders of magnitude, although we need to say that the lowest magnetic fields that we achieve in the voids actually represent the adiabatic decompressed field of the starting field of around $10^{-14}$ co-moving G at redshift 310. That means that the bulk of the volume is filled with a field stronger than $10^{-14}$ G (physical) at redshift zero. \\
Finally, we briefly discuss the panel on the bottom right of Fig.~\ref{fig:2dpdfs} where we show the magnetic field strength normalized to the flux-freezing limit. We can clearly see that the low-density regions are actually slightly steeper than the flux freezing limit, while the high-density regions follow very well the prediction from either Kazantsev theory (golden line) for the saturation regime of the turbulent dynamo or the prediction of \citet{Xu2020} for the saturation regime of a turbulent dynamo under gravitational collapse (purple line). Despite the fact that we have very good resolution and gravitational motions are very well resolved there is not a clear trend of which of these lines is more accurate. This just demonstrates that the situation for the turbulent dynamo in a realistic system is more complex due to the scatter of the simulation itself. Hence, neither of the scenarios can be clearly ruled out based on our simulation. We do note that if we average the distribution under gravitational contraction at higher redshift we find a slope that is more consistent with the prediction of \citet{Xu2020} yielding a fitted slope of around $-0.1$. This is consistent with our earlier predictions in \citet{SteinwandelClusterpaper} for the lower-resolution versions of our high-res adiabatic galaxy cluster simulations in this work.

\subsection{Drivers of the rate of change of the magnetic field}

However, instead of repeating the analysis on our older, lower-resolution runs we want to stress that in this work we aimed for a more rigorous investigation of magnetic field amplification which we plan to extend in future studies. In Fig.~\ref{fig:bb} we aim to disentangle the magnetic field amplification (suppression) due to shear and compression. Hereby, we focus on aspects of the whole simulation domain which allows us to unravel that adiabatic compression at shock fronts in the low redshift Universe is heavily responsible for the high positive rate of change of the magnetic field, while the shear appears to have a negative impact on these compressed regions. While the former is quite intuitive, the latter is harder to disentangle since the term $\hat{\mathbf{b}} \hat{\mathbf{b}}:\nabla \mathbf{u}$ has nine components. However, since the feature in the cold front itself is rather sharp it indicates that the magnetic field direction is perpendicular to the bulk flow which can explain the change in sign for the shearing component in the compressed region. This is actually a similar situation to the one reported by \citet{Komarov2016} in their Fig. 5. It is interesting to point out that while we find that the compressive mode of amplification is dominant at the shocks, these represent only a very small volume of the ICM where the ICM gas is behaving compressible. In the vast majority of the volume, the adiabatic term plays a minor role and shear is dominating the amplification in a volume-weighted picture.  

\subsection{Large scale magnetic fields in the ICM}

\begin{figure*}
    \centering
    \includegraphics[width=\textwidth]{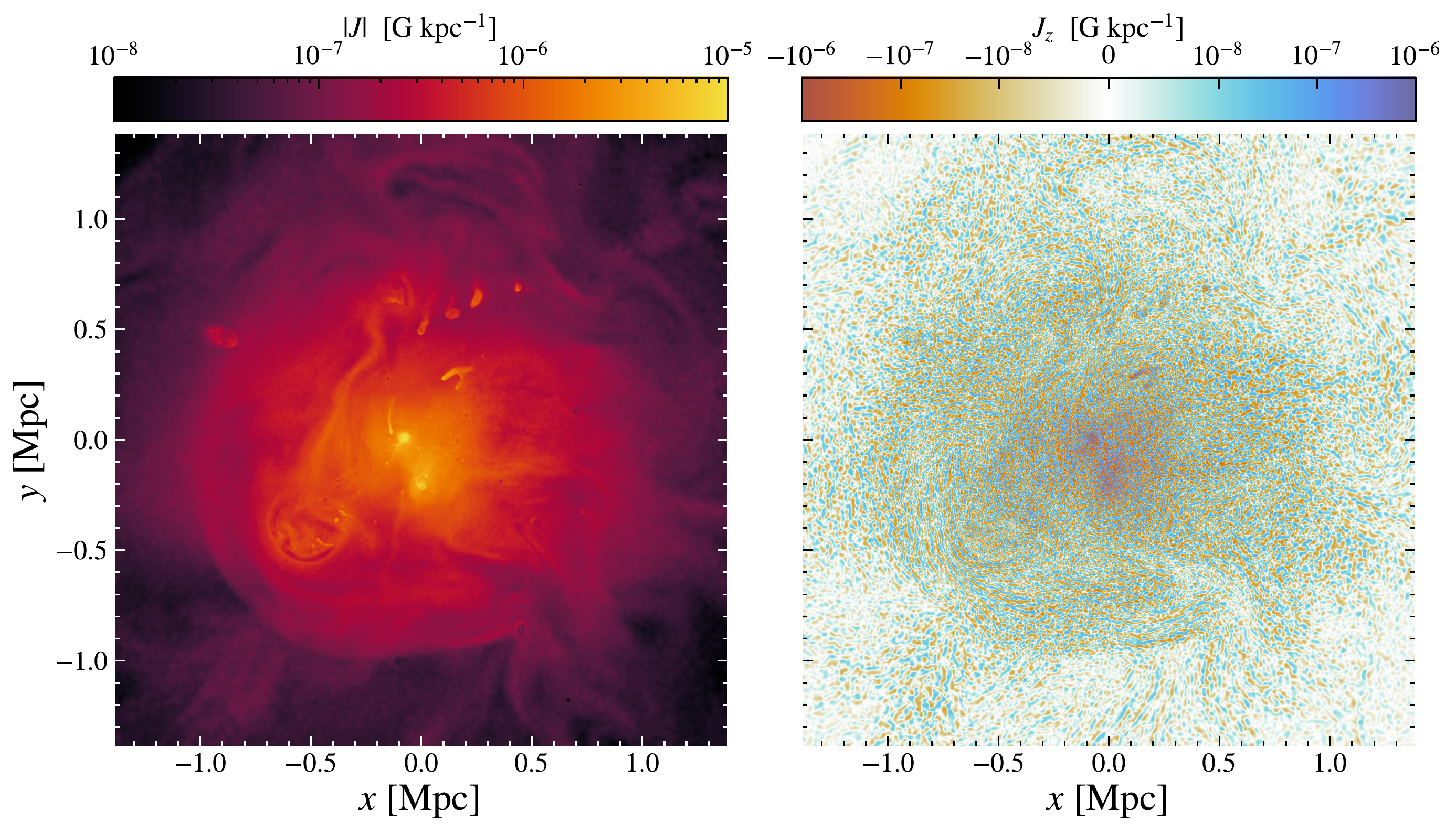}
    \caption{Left: Absolute value of $\mathbf{J} = \nabla \times \mathbf{B}$ showing detailed small scale structure. Right: $J_{z} = \partial_\mathrm{x} B_\mathrm{y} - \partial_\mathrm{y} B_\mathrm{x}$. We find rapid changes in sign that indicate magnetic re-connection.}
    \label{fig:b_rotb}
\end{figure*}

Finally, yet importantly we want to discuss the implications of the fluctuations of the radial and angular field components as shown in Fig.~\ref{fig:b_structure} for B$_{r}$ (black), B$_{\theta}$ (red) and B$_{\phi}$ (green). We plot these twice as a function of radius, once normalized to the virial radius and once on a physical scale, which is the left and the right-hand panels of Fig.~\ref{fig:b_structure}. The lines in the plot represent the clusters' mean field. This is important to highlight since there is some confusion about the term mean field in the literature where some papers assume the quantity that we dubbed B$_\mathrm{rms}$ as the clusters mean-field, but the mean field is the actual mean of the single components of the vector of $\mathbf{B}$. We find rather lower values of around 0.1 $\mu$G in the ICM where radial and angular components fluctuate on a physical scale of around 10 kpc. This gives us a lower limit for the length scale on which magnetic fields are correlated. Additionally, we find that the mean field is essentially zero above a scale of 500 kpc, which gives us an upper limit for the magnetic field's correlation length. Comparing this length-scale to the size of magnetized structures in Fig.~\ref{fig:cluster_yz} this seems reasonable. We note that the mean radial and angular components of the field show trends to be anti-correlated with one another. This is not very surprising since a dynamo will convert radial field into the angular field and vice versa by definition. It is interesting to point out that the field is quite symmetric in all these components which we show in Fig.~\ref{fig:b_1dpdfs} that shows the one-dimensional PDF for the radial and angular components of the field. There seems to be an excess of the radial field around 0.2 $\mu$G but the difference is marginal and the lines are almost identical. What this likely means is that the structure formation process is dominant for ordering the field on Mpc-scales while the turbulent dynamo is amplifying it with a maximum correlation length of a few 100 kpc.   

Finally, we want to briefly discuss our findings for $\mathbf{J} = \nabla \times \mathbf{B}$ and its consequences for magnetic reconnection in the ICM. No dynamo will work without magnetic re-connection, as the complex interplay between amplification of the field due to stretching twisting and folding is limited by reconnection on smaller scales which will dissipate magnetic field energy and convert it into heat. This will actually set the dynamos' growth rate and determine its saturation level. However, in cosmological simulations of realistic systems dynamo properties are studied by various different groups as referenced above while re-connection properties are typically ignored. There are two reasons for this first. Re-connection in most MHD simulations (at least on our scale) typically relies on numerical diffusivity for the reconnection part which is not ideal. However, our simulation assumes a constant magnetic resistivity, which allows us to make a least some very basic statements about magnetic re-connection. For instance, we can look at the absolute value of $\mathbf{J}$ and its single components. We show an example of this in Fig.~\ref{fig:b_rotb}. While both of these reveal interesting sub-structure it is especially apparent from $J_\mathrm{z}$ that reconnection seems to appear on all scales in the ICM as $J_\mathrm{z}$ is rapidly switching signs. Obviously, a more targeted study is needed to disentangle to which degree this is resolved in our simulation, despite the fact that these structures seem to closely resemble the typical shapes of ``Plasmoids''. 

\section{Conclusion}

We presented the first results for a high-resolution simulation with magnetic fields in which we put a specific focus to study the ICMs magnetic field structure in greater detail and compare to previous simulations, including our own that achieved at least a factor of 10 lower mass and factor of around 4 lower spatial resolution. Our most important findings are: 

\begin{enumerate}

    \item We present a galaxy cluster simulation that resolves the Coulomb mean free path for a large fraction ($>80$ per cent) of the Lagrangian mass of the systems. This motivates future studies at our resolution with either some kinetic aware theory (e.g. Braginskii-MHD) or some more sophisticated kinetic treatment in fully cosmological simulations of the magnetized ICM.
    \item The total energy budget of the magnetic field saturates at a level of $\sim$ 10 per cent of the turbulent kinetic energy within the cluster. This is comparable with the saturation level in turbulent box simulations in the high magnetic Prandtl number regime. This is encouraging because, on the one hand, this makes the approximation of the ICM as a ``turbulent box'' a valid one in terms of the energetics. On the other hand, it is encouraging that we achieve the saturation level of typical dynamo simulations in the turbulent-driven regime in a more realistic simulation in a fully cosmological assembly scenario.
    \item The mean magnetic field strength peaks at around redshift 1.5 at a value of 10 $\mu$G, while the field at redshift zero is lower by a factor of around two and saturates at a value of 5 $\mu$G. 
    \item The radial trend of the magnetic fields towards the center predicts magnetic fields that are higher than the ones obtained by RM-measurements of nearby clusters by a factor of two to three. 
    \item We investigated the origin of the large magnetic fields and find that the underlying issue is driven by the steep entropy profile that we find in the cluster center, despite the fact that our simulation is adiabatic.
    \item In most regions within the cluster, turbulence is sub-sonic and trans-alfv\'{e}nic. However, this changes in the outskirts where the field is weak and turbulence is mostly super-alfv\'{e}nic and sub-sonic, apart from the regions in which the strong internal shocks drive the ICM in a mildly super-sonic regime. The highest Mach numbers we find are related to the accretion shock that is located at around 2.3 R$_\mathrm{vir}$.
    \item The small-scale magnetic field is correlated on a length-scale with from 10 kpc to a few 100 kpc.
    \item The mean radial and angular components as a function of the radius are anti-correlated which demonstrates the dynamo origin of the magnetic fields in our simulation.
    \item We investigated the spatial structure of the current $\mathbf{J}$ and find evidence for magnetic reconnection on all spatial scales within the ICM, motivating further studies on magnetic reconnection in the ICM. 
\end{enumerate}

Finally, we note that this is the first fully cosmological MHD simulation of a massive galaxy cluster with a mass of around $2 \times 10^{15}$ M$_{\odot}$ with $\sim 10^{9}$ resolution elements in the virial radius at redshift zero. This results in a mass resolution of $4 \times 10^{5}$ M$_{\odot}$ making this (to our knowledge) the highest resolution galaxy cluster simulation of a system beyond $10^{15}$ M$_{\odot}$ to date.  

\section*{Data Availability}
The data will be made available based on reasonable request to the corresponding author.
\acknowledgments

\vspace{5mm}
UPS is grateful to Jonathan Squire, Lorenzo Sironi, and Aaron Tran for dicussion on Fig.~\ref{fig:bb}. UPS thanks Amitava Bhattacharjee, Greg L. Bryan, Blakesley B. Burkhart, Drummond B. Fielding, R\"{u}diger Pakmor, Rachel S. Somerville, Romain Teyssier as well as the members of CCAs Galaxy Formation and Plasma Physics groups for insightful and useful discussion. We thank Lucy Redding-Ikkanda for help with generating Fig.~\ref{fig:dynamo}. UPS is supported by the Simons Foundation through a Flatiron Research Fellowship (FRF) at the Center for Computational Astrophysics. The Flatiron Institute is supported by the Simons Foundation. UPS acknowledges computing time provided by the resources at the Flatiron Institute on the cluster rusty. UPS acknowledges the computing time provided by the Leibniz Rechenzentrum (LRZ) of the Bayerische Akademie der Wissenschaften on the machine SuperMUC-NG (pn72bu). UPS is acknowledging the computing time provided by c2pap (pr27mi).
KD, LMB and TM acknowledge support by the COMPLEX project from the European Research Council (ERC) under the European Union’s Horizon 2020 research and innovation program grant agreement ERC-2019-AdG 882679.
LMB acknowledges support by the Deutsche Forschungsgemeinschaft (DFG, German Research Foundation) under Germany´s Excellence Strategy – EXC 2094 – 390783311.
LMB acknowledges the computing time provided by c2pap under the project pn36ze.

\software{We use the cosmological simulation code \textsc{gadget3} \citep[][]{Springel2005, Dolag2009, Beck2016} to run the simulations and use the language \textsc{julia} \citep[][]{Bezanson2014} to perform the analysis\footnote{\url{https://docs.julialang.org}} based on the packages that can be found here: \url{https://github.com/LudwigBoess}}

\clearpage
\bibliography{sample63}{}
\bibliographystyle{aasjournal}

\end{document}